\documentclass[a4paper,12pt]{article}
\usepackage[a4paper,text={16.8cm,22.4cm}]{geometry}
\usepackage{amsmath,amssymb,bm,psfrag,graphicx,color}
\usepackage{multirow,cite} 
\allowdisplaybreaks 
\def\eps{\epsilon}
\def\mutilde{\tilde \mu}
\def\calO{\mathcal{O}}
\def\dilog{\hspace{0.4mm}\mbox{Li}_2}
\def\trilog{\hspace{0.4mm}\mbox{Li}_3}
\def\nielsen{\hspace{0.6mm}\mbox{S}_{1,2}}

\begin{document}

\begin{titlepage}

\begin{flushright}
October 1, 2012
\end{flushright}

\vspace{0.2cm}
\begin{center}
\Large\bf
{\boldmath
NNLL  Resummation for Jet Broadening
\unboldmath}
\end{center}

\vspace{0.2cm}
\begin{center}
Thomas Becher and Guido Bell\\
\vspace{0.4cm}
{\sl 
Albert Einstein Center for Fundamental Physics\\
Institut f\"ur Theoretische Physik, Universit\"at Bern\\
Sidlerstrasse 5, CH--3012 Bern, Switzerland}
\end{center}

\vspace{0.2cm}
\begin{abstract}
\vspace{0.2cm}
\noindent 
The resummation for the event-shape variable jet broadening is extended to next-to-next-to-leading logarithmic accuracy by computing the relevant jet and soft functions at one-loop order and the collinear anomaly to two-loop accuracy. The anomaly coefficient is extracted from the  soft function and expressed in terms of polylogarithmic as well as elliptic functions. With our results, the uncertainty on jet-broadening distributions is reduced significantly, which should allow for a precise determination of the strong coupling constant from the existing experimental data and provide a consistency check on the extraction of $\alpha_s$ from higher-log resummations of thrust. 
\end{abstract}
\vfil

\end{titlepage}

\tableofcontents

\newpage

\section{Introduction}

Event-shape variables characterize geometrical properties of final states observed in high-energy collisions. They are the simplest class of observables which probe properties beyond the total cross section and have been measured with high accuracy at LEP and other $e^+e^-$ machines. Due to their inclusive nature, they have little sensitivity to hadronization effects and can be computed perturbatively. The canonical event shape is thrust $T$. To evaluate the thrust of an event, one first determines the thrust axis $\vec{n}_T$, which is the direction of maximum momentum flow. The thrust is then defined as the fraction of particle momentum flowing along the thrust axis. The variable we consider here is the jet broadening, which measures the momentum flow orthogonal to the thrust axis. The total jet broadening is defined as
\cite{Rakow:1981qn,Ellis:1986ig}
\begin{equation*}
   b_T = \frac{1}{2} \sum_{i} |{\vec{p}_i}^\perp| 
   = \frac{1}{2} \sum_{i} |\vec{p}_i \times \vec{n}_T| \,.
\end{equation*}
By using the thrust axis to split the event into left and right hemispheres, one can define left and right broadenings $b_L$ and $b_R$. The total broadening is the sum of these, $b_T=b_L+b_R$, and the wide broadening is defined as the maximum of the two, $b_W={\rm max}(b_L,b_R)$. One often normalizes the broadening to the total momentum flow which, for massless particles, is equal to the center-of-mass energy $Q$ of the collision. We use capital letters to denote the normalized values, for example $B_T = b_T/Q$.

For events consisting of two low-mass jets recoiling against each other, the thrust is near the maximum value $T=1$, while the broadening is very small. In this region, higher-order corrections are enhanced by large logarithms which need to be resummed to all orders to obtain reliable predictions. For total broadening the leading logarithms have the form $\alpha_s^n \ln^{2n}(B_T)$. For thrust the resummation has been performed to next-to-next-to-next-to-leading logarithmic (N$^3$LL) accuracy \cite{Becher:2008cf}, while for broadening only NLL resummation is available \cite{Dokshitzer:1998kz}. For both observables, the next-to-next-to-leading order (NNLO) perturbative corrections are known \cite{GehrmannDeRidder:2007hr,Weinzierl:2009ms}. For thrust, combining the fixed-order corrections with the resummed result, an accurate value for the strong coupling constant has been determined from a global fit to the available experimental data \cite{Abbate:2010xh}. This yields one of the most precise determinations of $\alpha_s$ currently available. Interestingly, the value comes out almost $4\sigma$ below the world average \cite{Beringer:1900zz}. It is important to validate this result using other event-shape variables. To this end, we extend in this paper the resummation for the broadening distributions to NNLL accuracy.

So far, the only other event-shape variable for which the resummation has been carried out to higher accuracy is the heavy jet mass \cite{Chien:2010kc}.  In the two-jet limit it is, however, closely related to thrust. In contrast, the jet broadening is not only literally orthogonal, but also its theoretical description in the two-jet region differs in important aspects. Both observables receive contributions from soft emissions and from radiation collinear to the jets, but the momenta scale differently in the two cases.
For broadening, the transverse momentum of all emissions is of the same order as the broadening, $|p_s^\perp| \sim |p_c^\perp| \sim b_T$. In contrast, the transverse momenta of  soft particles contributing to thrust are power suppressed, $(p^\perp_s/ p^\perp_c)^2 \sim (1-T)$ near the end-point $T=1$. Because of this scaling, broadening is sensitive to soft recoil effects, while thrust is not. These recoil effects modify the higher-log resummation \cite{Dokshitzer:1998kz} and were missed in an early analysis \cite{Catani:1992jc}.

For the above reason, also the description of the two event shapes in Soft-Collinear Effective Theory (SCET) \cite{Bauer:2000yr,Bauer:2001yt,Beneke:2002ph} differs. The effective theory relevant for thrust involves soft particles with power suppressed momenta, which are also called ultra-soft, while the components of the soft momenta are unsuppressed for broadening. The versions of SCET relevant in the two cases are sometimes called SCET$_{\rm I}$ and SCET$_{\rm II}$. While the effective-theory analysis of SCET$_{\rm I}$ observables is well understood, this was not the case for SCET$_{\rm II}$. One problem was that the loop integrals in the effective theory were not always well-defined in dimensional regularization, another that renormalization group (RG) invariance required the low-energy effective theory to depend on the hard momentum scale, which corresponds to the center-of-mass energy $Q$. This dependence arises naturally in SCET$_{\rm I}$ from the power suppressed momentum of the soft modes, but is naively absent in SCET$_{\rm II}$. Both issues are now understood. The additional singularities can be regularized with an analytic regulator. The individual soft and collinear contributions then suffer from singularities in this regulator, which cancel once the contributions are combined. However, the presence of the regulator breaks a rescaling symmetry which is not recovered when the regulator is removed. This effect is called the collinear factorization anomaly \cite{Becher:2010tm}. It induces $Q$-dependence in the low-energy theory. Consistency conditions can be used to show that this $Q$-dependence exponentiates \cite{Becher:2010tm,Chiu:2007dg}. Instead of deriving the exponentiation from consistency conditions, one can also obtain it from an RG framework, the so-called rapidity RG \cite{Chiu:2011qc,Chiu:2012ir}. The associated framework is special in that one does not evolve from a higher to a lower virtuality, and that there is no physical coupling constant associated with the evolution. Nevertheless, using this formalism one arrives at the same exponentiated $Q$-dependence. The early papers on SCET$_{\rm II}$ processes used versions of analytic regularization which break important properties of the effective theory such as gauge invariance and the eikonal structure of soft emissions. We have recently shown that only the phase-space integrals suffer from the additional divergences and that the regulator can be introduced in such a way that the factorization structure of the effective theory is maintained \cite{Becher:2011dz}. The recent work \cite{Gehrmann:2012ze} and our paper provide the first examples of two-loop computations in SCET for anomalous observables. The simple form and the good properties of our regulator were crucial to make these computations feasible.  

Based on the new formalism, an all-order factorization theorem for the broadening distributions was derived in \cite{Becher:2011pf}. This  allows for higher-log resummations, but in contrast to thrust the available perturbative input was only sufficient for NLL accuracy. In this paper, we derive the missing ingredients for NNLL resummation. These include the one-loop jet and soft functions as well as the two-loop anomaly exponent. The necessary computations are quite involved. In addition to polylogarithmic functions the results also contain elliptic integrals. As a check, we have independently performed the computations analytically and numerically. At small values of the broadening, our computation fully predicts the broadening distributions at ${\mathcal O}(\alpha_s^2)$, and also all logarithmic terms at ${\mathcal O}(\alpha_s^3)$. By comparing to numerical fixed-order results from event generators, we verify that we indeed obtain the correct behavior in the two-jet limit. Finally, we  present numerical predictions for the resummed rates and discuss scale uncertainties and scheme dependence.

Our paper is organized as follows. We start by discussing the factorization theorem for jet broadening and defining the relevant jet and soft functions. In Section \ref{sec:oneloop} we perform the one-loop computations of the jet and soft functions. The main result of our paper is the computation of the two-loop anomaly coefficient, which is described in Section \ref{sec:twoloop}. The two sections concerning the perturbative computations are necessarily technical. The reader only interested in the results can find these in Sections \ref{sec:W} and \ref{sec:anomaly}. We then derive the NNLL resummed result in Section \ref{sec:resummation} and present predictions for the broadening distributions at LEP energies. In Section \ref{sec:fixed} we compare the perturbative expansion of the resummed result to numerical results of fixed-order event generators. The necessary expansion coefficients are listed in Appendix \ref{sec:expcoeff}. The numerical evaluation of the jet and soft diagrams is discussed in Appendix \ref{sec:numerical}, and the relevant two-loop matrix elements are given in Appendix \ref{sec:softmatrix}. Analytic results for the individual diagrams, both in momentum and in Fourier-Laplace space, can be found in Appendix \ref{sec:diagrams}.

\section{Factorization theorem for jet broadening\label{sec:fac}}

Collider events with low values of the broadening consist of two nearly massless jets recoiling against each other, accompanied by soft radiation. The relevant factorization theorem has the form \cite{Chiu:2011qc,Becher:2011pf}
\begin{equation}
\begin{aligned}\label{naivefac}
   \frac{1}{\sigma_0}\,\frac{d^2\sigma}{db_L\,db_R} 
   &= H(Q^2,\mu) \int\!db_L^s \int\!db_R^s \int\!d^{d-2}p^\perp_L \int\!d^{d-2}p^\perp_R \\
   &\quad\times {\cal J}_L(b_L-b_L^s,p_L^\perp, \mu)\, {\cal J}_R(b_R-b_R^s,p_R^\perp,\mu)\, 
    {\cal S}(b_L^s,b_R^s,-p_L^\perp, -p_R^\perp,\mu) \,.
\end{aligned}
\end{equation}
The hard function $H(Q^2,\mu)$ describes the production of the quark pair at large momentum transfer, before collinear and soft emissions.  It is the square of the quark vector form factor $H(Q^2,\mu)=|C_V(-Q^2-i\varepsilon,\mu)|^2$, which is known to three-loop accuracy \cite{Baikov:2009bg,Gehrmann:2010tu}. What distinguishes jet broadening from other dijet event shapes, such as thrust or jet masses, is that the transverse momentum of the soft emissions is of the same size as the one of the collinear partons. As a consequence, the jets recoil against the soft radiation, and the factorization theorem involves an integral over the transverse momentum. The broadening is the sum of the soft and collinear broadenings, and the transverse momentum of the collinear radiation is equal and opposite to the transverse momentum of the soft radiation.

The broadening jet function for the left-moving collinear partons is defined as
\begin{align}\label{jet:definition}
   \frac{\pi}{2}\,({n\!\!\!/})_{\alpha\beta}\,{\cal J}_L(b_L,p_L^\perp,\mu) 
   &= \sum\hspace{-0.75cm}\int\limits_{X, {\rm reg.}}\,\,(2\pi)^d\,\delta(\bar{n}\cdot p_X-Q)\, 
    \delta^{d-2}(p_X^\perp-p_L^\perp) \nonumber\\[-2mm]
   &\quad\times \delta\Big(b_L-\frac{1}{2} \sum_{i\in X} |p_i^\perp|\Big)\,   
    \langle 0|\chi_\alpha(0)|X\rangle\,\langle X|\bar{\chi}_\beta(0)|0\rangle \,.
\end{align}
This definition involves the reference four-vectors $n^\mu = (1,\vec{n}_T)$ and $\bar{n}^\mu = (1,-\vec{n}_T)$ along the thrust direction. The field $\chi(x)$ is the collinear quark field of SCET.  The collinear SCET Lagrangian is equivalent to the ordinary QCD Lagrangian and the field $\chi(x)$ can be identified with $\chi(x)=\frac{n\!\!\!/\bar n\!\!\!/}{4}\,W^\dagger(x)\,\psi(x)$, where $\psi(x)$ is the QCD quark field and $W(x)$ a straight Wilson line along the $\bar{n}^\mu$ direction from $-\infty$ to $x$ (see e.g.\ \cite{Becher:2009qa} for more details). Also the soft SCET Lagrangian is equivalent to the usual QCD Lagrangian and in all our computations, we will work with the standard QCD Lagrangian and Feynman rules.

Beyond tree level, a naive definition of the jet and soft functions leads to integrals that are not fully regularized by dimensional regularization, because they suffer from rapidity divergences. The sum over intermediate states in the above definition contains a prescription to regularize these divergences. It consists in a modification of the phase-space integrals and amounts to the replacement \cite{Becher:2011dz}
\begin{equation}\label{regulator}
\int \!d^dk \,  \delta(k^2) \theta(k^0)\;\; \to\;\; 
\int\!d^dk \left(\frac{\nu_+}{k_+}\right)^\alpha \,  \delta(k^2) \theta(k^0) \,,
\end{equation}
where $k_+ = n \cdot k$. As the rapidity divergences only occur in the phase-space integrals of the additional emissions, we find it convenient to use the standard measure for the tree-level phase space of the $q\bar q$ pair. The jet function ${\cal J}_R$ for the right-moving collinear partons is obtained by exchanging $n^\mu\leftrightarrow\bar{n}^\mu$ in the definition (\ref{jet:definition}), but keeping the same phase-space regulator $(\nu_+/k_+)^\alpha$. Also the broadening soft function, 
\begin{align}\label{soft:definition}
   {\cal S}(b_L,b_R,p_L^\perp,p_R^\perp,\mu) 
   &= \sum\hspace{-1.10cm}\int\limits_{X_L,X_R, {\rm reg.}}
	\!\!\!\delta^{d-2}(p_{X_L}^\perp-p_L^\perp)\, 
    \delta^{d-2}(p_{X_R}^\perp-p_R^\perp) \\[-2mm]
   &\quad\times \delta\Big(b_L-\frac{1}{2} \sum_{i\in X_L} |p_{L,i}^\perp|\Big)\, 
    \delta\Big(b_R-\frac{1}{2} \sum_{j\in X_R} |p_{R,j}^\perp|\Big) 
    \left| \langle X_L\,X_R| S_n^\dagger(0)\,S_{\bar{n}}(0) |0\rangle \right|^2 ,\nonumber
\end{align}
is regularized in the same way. After the convolution of the jet and soft functions has been carried out, one can take the limit $\alpha \to 0$. The individual functions suffer from divergences in this limit, which cancel in the convolution.

The jet and soft functions, as well as the broadening cross section itself, are distribution-valued both in the broadening and in the transverse momentum variables. To work with regular functions and to turn the convolutions in (\ref{naivefac}) into a product, it is advantageous to perform a Laplace transform of the cross section
\begin{equation}
   \frac{d^2\sigma}{d\tau_L\,d\tau_R} 
   = \int_0^\infty\!db_L\,e^{-\tau_L b_L} \int_0^\infty\!db_R\,e^{-\tau_R b_R}\,
    \frac{d^2\sigma}{db_L\,db_R} \,
\end{equation}
and to work in impact parameter space rather than transverse momentum space. To this end, we Laplace and Fourier transform the jet and soft functions. For the left jet function, for example, we define
\begin{equation}
   \widetilde{\cal J}_L(\tau_L,x_L^\perp,\mu)
   = \int_0^\infty\!db_L\,e^{-\tau_L b_L} \int\frac{d^{d-2}p_L^\perp}{(2\pi)^{d-2}}\,
    e^{-ip_L^\perp\cdot x_L^\perp}\,{\cal J}_L(b_L,p_L^\perp,\mu) \,,
\end{equation}
and analogously, for the right jet function and the soft function. One can further simplify the functions by averaging over the azimuthal angle, and define
\begin{equation}
\overline{\cal S}(\tau_L,\tau_R,z_L, z_R,\mu) 
   = \frac{1}{{\cal N}^2} \int\!d\Omega^L_{d-2} \int\!d\Omega^R_{d-2}\, 
    \widetilde{\cal S}(\tau_L,\tau_R,x_L^\perp,x_R^\perp,\mu)\,,
\end{equation}    
where we have introduced the dimensionless variables $z_{L,R}=2|x_{L,R}^\perp|/\tau_{L,R}$. The normalization factor
\begin{equation}
   {\cal N} = \frac{\Omega_{d-2}}{(2\pi)^{d-2}}
   = \frac{2}{(4\pi)^{1-\epsilon}\,\Gamma(1-\epsilon)}\,
\end{equation}
ensures that $\overline{\cal S}=1$ at lowest order, for any value of the space-time dimension $d=4-2\epsilon$. Introducing also
\begin{equation}
 \overline{\cal J}_{L,R}(\tau,z,\mu) 
   = {\cal N}\,(2\pi)^{d-2}\,\frac{\tau}{2} \left(\frac{\tau z}{2}\right)^{d-3}\, 
    \widetilde{\cal J}_{L,R}(\tau,x^\perp,\mu) \,,
\end{equation}
the factorization theorem in Laplace-Fourier space takes the simple form
\begin{equation}\label{factfinal}
   \frac{1}{\sigma_0}\,\frac{d^2\sigma}{d\tau_L\,d\tau_R} 
   = H(Q^2,\mu) \int_0^\infty\!dz_L \int_0^\infty\!dz_R\,
    \overline {\cal J}_L(\tau_L,z_L,\mu)\,\overline{\cal J}_R(\tau_R,z_R,\mu)\, 
    \overline{\cal S}(\tau_L,\tau_R,z_L, z_R,\mu) \,.
\end{equation}
As discussed above, the naive factorization theorems (\ref{naivefac}) and (\ref{factfinal}) do not achieve complete factorization, since the additional regulator induces implicit $Q$-dependence.
The main result of our previous paper \cite{Becher:2011pf} was to derive the all-order form of this $Q$-dependence and to show that the product of jet and soft functions can be written as
\begin{multline}\label{JJSfact}
   \overline {\cal J}_L(\tau_L,z_L,\mu)\,\overline{\cal J}_R(\tau_R,z_R,\mu)\, 
    \overline{\cal S}(\tau_L,\tau_R,z_L, z_R,\mu) \\
    =\exp\left[-F_B(\tau_L,z_L,\mu) \ln\big(Q^2\bar\tau_L^2\big) -F_B(\tau_R,z_R,\mu)
    \ln\big(Q^2\bar\tau_R^2\big)\right] \, W(\tau_L,\tau_R,z_L,z_R,\mu) \,,
\end{multline}
where the anomaly exponent $F_B(\tau_L,z_L,\mu)$ and the remainder function $W(\tau_L,\tau_R,z_L,z_R,\mu)$ are independent of the large scale $Q$. The results (\ref{factfinal}) and (\ref{JJSfact}) form the basis of the all-order resummation of large logarithms. There are two sources of dependence on the large scale $Q$: the hard function $ H(Q^2,\mu)$ and the anomaly in (\ref{JJSfact}). The large logarithms in the hard function can be resummed by solving the RG equation for this function and evolving from a high scale $\mu_h \sim Q$ to a low scale $\mu \sim 1/\tau_{L/R} \sim 1/b_{L/R}$. The logarithms associated with the anomaly are exponentiated in (\ref{JJSfact}). To achieve NNLL accuracy, we need to evaluate the function $W(\tau_L,\tau_R,z_L,z_R,\mu)$ to one-loop accuracy and the anomaly exponent $F_B(\tau,z,\mu)$ to two-loop accuracy. The computation of the jet and soft functions to one-loop order will be discussed in the next section. The two-loop anomaly exponent can be extracted from a computation of the divergences of the two-loop soft function in the analytic regulator, as discussed in detail in Section \ref{sec:anomaly}.

\section{Jet and soft functions at one-loop order\label{sec:oneloop}}

\subsection{Soft function}

\begin{figure}[t!]
\begin{center}
\begin{tabular}{ccccccc}
\includegraphics[width=0.19\textwidth]{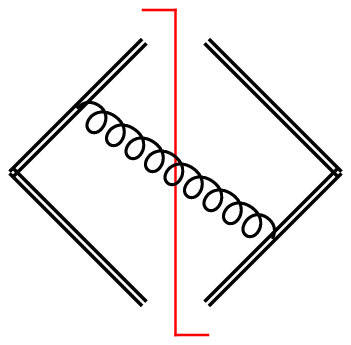} & & 
\includegraphics[width=0.19\textwidth]{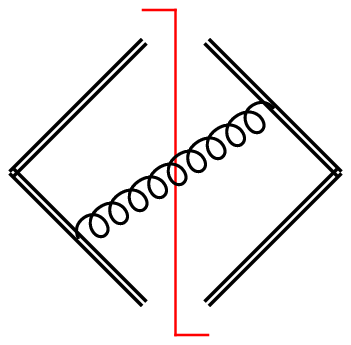} & &
\includegraphics[width=0.19\textwidth]{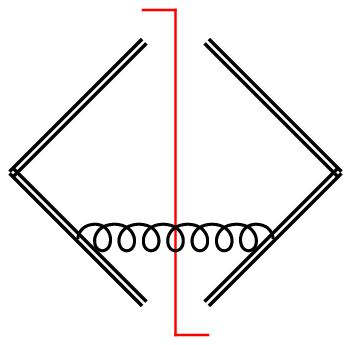} & & 
\includegraphics[width=0.19\textwidth]{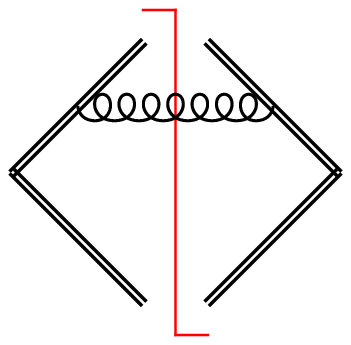}
\end{tabular}
\end{center}
\vspace{-0.5cm}
\caption{\label{soft:diagrams}
Next-to-leading corrections to the soft function. The one-loop virtual diagrams are scaleless and vanish.}
\end{figure}

We first compute the one-loop soft function, which is simple since there is at most one soft particle in the final state at the given order. The one-loop virtual corrections are scaleless and vanish in dimensional regularization. The real emission diagrams are shown in Figure \ref{soft:diagrams}. As the last two diagrams vanish and the first two diagrams give identical contributions, one is left with a single diagram at this order. The result for the one-loop soft function has already been given in our previous work in a slightly different regularization prescription \cite{Becher:2011pf}. Setting $\beta=0$ and $\nu_1^2=\nu_+ Q$, one can easily translate these expressions to the current scheme. In broadening-momentum space the result takes the form
\begin{align}
   {\cal S}(b_L,b_R,p_L^\perp, p_R^\perp) 
   &= \delta(b_L)\,\delta(b_R)\,\delta^{d-2}(p_L^\perp)\,\delta^{d-2}(p_R^\perp)
	\\
	&\quad
	-\frac{\alpha_s C_F}{\pi^{2-\epsilon}} \left( \mu^2 e^{\gamma_E} \right)^\epsilon\,
    \frac{\nu_+^\alpha}{\alpha}\,
	\bigg\{ p_L^{-2-\alpha}\,
    \delta\Big(b_L-\frac{p_L}{2}\Big)\,\delta(b_R)\,\delta^{d-2}(p_R^\perp)
    - (L\leftrightarrow R) \bigg\},\nonumber
\end{align}
with $p_L=|p_L^\perp|$. After Laplace and Fourier transformation this yields
\begin{align}
   \overline{\cal S}(\tau_L,\tau_R,z_L, z_R) 
   &= 1 + \frac{\alpha_s C_F}{4\pi}\bigg\{
	\big( \mu^2 \bar \tau_L^2\big)^\eps\; \big(\nu_+ \bar \tau_L\big)^\alpha\;	
	 \bigg[\frac{4}{\alpha} 
		\bigg( \frac{1}{\eps} + 2 \ln z^L_+ \bigg)\nonumber
	- \frac{2}{\eps^2}  
\\
   &\hspace{2.6cm}
	+8\dilog\Big(\! -\frac{z^L_-}{z^L_+} \Big) 
	+ 4 \ln^2 z^L_+ + \frac{5\pi^2}{6}\bigg]
	-(L\leftrightarrow R)\bigg\}\,,
\end{align}
where we introduced the notation $z^L_\pm = (\sqrt{1+z_L^2}\pm1)/4$ and $\bar\tau_{L}=  \tau_L e^{\gamma_E}$.

\subsection{Jet function}
\label{sec:jet}

\begin{figure}[t!]
\begin{center}
\begin{tabular}{ccccccc}
\includegraphics[width=0.19\textwidth]{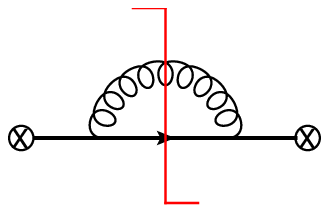} & & 
\includegraphics[width=0.19\textwidth]{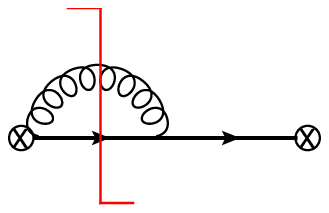} & &
\includegraphics[width=0.19\textwidth]{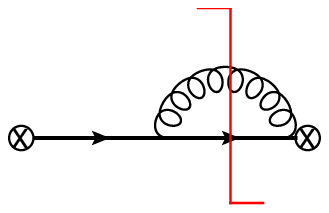} & & 
\includegraphics[width=0.19\textwidth]{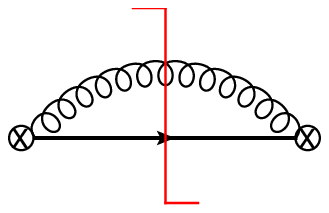}
\end{tabular}
\end{center}
\vspace{-0.5cm}
\caption{\label{jet:diagrams}
Next-to-leading corrections to the jet function. The one-loop virtual corrections are scaleless and vanish.}
\end{figure}

The calculation of the jet function is considerably more complicated, since there are now up to two collinear partons in the final state, and the integrals involve a non-trivial angle in the transverse-momentum plane. At tree level, the left and right jet functions are given by delta functions, 
${\cal J}_{L,R}^{(0)}(b,p^\perp)=\delta(b-\frac12 |p^\perp|)$, which after Laplace and Fourier transformation turn into
\begin{equation}\label{jet:tree}
   \overline{\cal J}^{(0)}_{L,R}(\tau, z) 
   = \frac{4^{\epsilon}\,\Gamma(2-2\epsilon)}{\Gamma^2(1-\epsilon)}\,
    \frac{z^{1-2\epsilon}}{\left(1+z^2\right)^{3/2-\epsilon}} \,.
\end{equation} 
In the following, we will drop the subscript $L,R$ of the tree-level jet function for convenience. At one-loop order, the left and right jet functions do not coincide anymore, since the regulator (\ref{regulator}) breaks the left-right symmetry, which is recovered only when the jet and soft functions are put together. The one-loop virtual corrections are again scaleless and vanish in dimensional regularization. Among the real emission diagrams, the last diagram  in Figure \ref{jet:diagrams} vanishes and the second and third diagrams give identical contributions. The jet functions can thus be written in the form
\begin{equation}\label{jet:NLO}
   {\cal J}_{L,R}(b,p^\perp)  
   = {\cal J}^{(0)}(b,p^\perp) 
	+ {\cal J}_{L,R}^{(1a)}(b,p^\perp) 
	+ 2\,{\cal J}_{L,R}^{(1b)}(b,p^\perp) + \calO(\alpha_s^2)\,,
\end{equation} 
where the term with the superscript $(1a)$ refers to the self-energy contribution in the first diagram in Figure \ref{jet:diagrams}, and the term $(1b)$ to the Wilson-line contribution in the second diagram. In the remainder of this section, we explain the analytic calculation of these diagrams. We discuss the left jet function in detail, but will only summarize the results for the right jet function. As a check of our calculations, we computed the jet functions numerically in Laplace-Fourier space. The details of our numerical approach can be found in Appendix \ref{sec:numerical:jet}.

\subsubsection*{Self-energy diagram}

In the original broadening-momentum space the self-energy contribution to the left jet function becomes
\begin{align}\label{self}
   {\cal J}_{L}^{(1a)}(b,p^\perp) 
   &= \alpha_s C_F\,\frac{2^{1+2\eps}}{\pi^{2-2\eps}} \;(1-\eps)\;\mutilde^{2\eps} \int d^dq \; \delta(q^2)\,\theta(q^0)\;
	\int d^dk \,\left(\frac{\nu_+}{k_+}\right)^\alpha \delta(k^2)\,\theta(k^0)
\nonumber\\
   &\quad\times\,
    \frac{k_-}{2 k\cdot q}\;\,
    \delta(Q-q_--k_-)\;
    \delta^{d-2}(p^\perp-q^\perp-k^\perp)\;
    \delta\Big(b-\frac12 |q^\perp| - \frac12 |k^\perp|\Big)\,,
\end{align}
where $q$ and $k$ are the quark and gluon momenta, respectively, and $ \mutilde^2 = \mu^2 e^{\gamma_E}/(4\pi)$. Notice that we did not impose an additional regulator on the phase space of the quark, since we work in a scheme where the tree-level phase space remains untouched. We now use the first delta function in the second line to perform the $k_-$-integration, the second delta function for $k^\perp$ and the on-shell conditions for the integrations over the plus components. We then arrive at
\begin{align}
   {\cal J}_{L}^{(1a)}(b,p^\perp) 
   &= \alpha_s C_F\,\frac{2^{-1+2\eps}}{\pi^{2-2\eps}} \;(1-\eps)\;\mutilde^{2\eps} 
   \int_0^Q \!dq_- \;  \int_0^\infty \!dq\; q^{d-3}\;
   \int_{-1}^1 \! d\cos\theta\; \sin^{d-5}\theta\;
   \int d\Omega_{d-3}
\nonumber\\
   &\quad\times\,
    \left(\frac{(Q-q_-)\nu_+}{p^2+q^2-2p q \cos\theta}\right)^\alpha
    \frac{Q-q_-}{q_-^2 p^2 +Q^2 q^2 - 2q_- Q p q \cos \theta}
\nonumber\\
   &\quad\times\,
    \delta\Big(b-\frac{1}{2}\,q - \frac12 \sqrt{p^2+q^2-2p q \cos\theta}\Big)\,,
\end{align}
where $\theta$ denotes the angle between $\vec{q}^\perp$ and $\vec{p}^\perp$, and we abbreviate $q=|q^\perp|$ and $p=|p^\perp|$.  Next, we use the remaining delta function to perform the integration over $\cos\theta$. It is furthermore convenient to introduce dimensionless variables via $q_- = Q \eta$, $q = b \xi$ and $y = p/2 b$. This gives
\begin{align}\label{jet:se:bp}
   {\cal J}_{L}^{(1a)}(b,p^\perp) 
   &= \frac{\alpha_s C_F}{2\pi^{1-\eps}}\,
   \left(\frac{\mu^2e^{\gamma_E}}{b^2}\right)^\eps \;\frac{(1-\eps)}{b}\;\Omega_{d-3}
   \;\left(\frac{\nu_+Q}{b^2}\right)^\alpha (1-y^2)^{-1-\eps} \;I_L^{(1a)}(y),
\end{align}
with the two-dimensional integral
\begin{align}
   I_L^{(1a)}(y) &= \frac{\sqrt{1-y^2}}{\pi} y^{2\eps}
   	\int_0^1 \!d\eta \,(1-\eta)^{1+\alpha}\!
	\int_{1-y}^{1+y} \!\!\!d\xi\;
	\frac{\xi (2-\xi)^{1-2\alpha} (1+y-\xi)^{-\frac12-\eps}(\xi-1+y)^{-\frac12-\eps}}
	{(\xi-2y\eta)^2+4\eta(1-y)(1+y-\xi)}\,.
\end{align}
Note that we factored out a distribution in (\ref{jet:se:bp}), that is divergent in the collinear limit $y\to1$.

As the above integral is well-defined in the limit $\alpha\to0$, we may set the analytic regulator to zero. We then integrate over $\eta$ and symmetrize the remaining $\xi$-integral. Substituting $\xi=1+y\sqrt{u}$ leads to the intermediate result
\begin{align}
   I_L^{(1a)}(y) &= \frac{1}{4\pi y} \,\int_0^1 du\;\,
	\frac{(1 - u y^2)}{\sqrt{u}}\;(1-u)^{-1-\eps}\,
	\bigg\{ \arctan \bigg( \frac{\sqrt{u}+y}{\sqrt{(1-y^2)(1-u)}} \bigg)
		- (y\to-y) \bigg\}\,.
\end{align}
We next perform a partial integration and rewrite the resulting hypergeometric functions with their integral representations. After inverting the order of the integrations, the integral can be solved in closed form and we find
\begin{align}
   I_L^{(1a)}(y) &= \frac{4^{-1+\eps}}{\sqrt{1-y^2}} \frac{\Gamma(1-2\eps)}{(1-\eps)\Gamma^2(1-\eps)}
\nonumber\\
   &\quad\times\,
	\bigg\{ 2 (1 - \eps) (1 - y^2) \,
	{}_3F_2\bigg(\frac12, 1, \frac12 - \eps; \frac32, 1 - \eps; - \frac{y^2}{1 - y^2} \bigg)
\nonumber\\
   &\hspace{1.1cm}
	+ y^2(1 - 2 \eps)\,{}_3F_2\bigg(\frac12, 1, \frac32 - \eps; \frac32, 2 - \eps;- \frac{y^2}{1-y^2}\bigg)\, \bigg\}\,.
\end{align}
In the collinear limit $y\to1$, we derive 
\begin{align}
   I_L^{(1a)}(y\to 1) &\simeq \frac{\Gamma(1 -\eps) \Gamma(2 - \eps)}{4^\eps\,\Gamma(3 - 2 \eps)}
	= \frac12 + (1 - \ln 2) \eps + \calO(\eps^2)\,.
\end{align}
As the prefactor of the integral in (\ref{jet:se:bp}) diverges only for $y\to 1$, the higher-order term is needed only in that limit. We may thus expand the exact expression for the integral as
\begin{align}
   I_L^{(1a)}(y) &= \frac12 - \frac12 \sqrt{1 - y^2} - \frac{1 - y^2}{4y} \ln\bigg(\frac{1 - y}{1 + y}\bigg)
	+ (1 - \ln 2) f(y) \,\eps + \calO(\eps^2)\,,
\end{align}
where the precise form of the auxiliary function $f(y)$ is not needed, but only its value at $f(1)=1$. After taking the Laplace and Fourier transform we finally obtain
\begin{align}\label{jet:se:tauz}
   \overline{\cal J}_{L}^{(1a)}(\tau, z) 
   &=    \overline{\cal J}^{(0)}(\tau, z) \,
	\frac{\alpha_s C_F}{4\pi}\;\big( \mu^2 \bar \tau^2\big)^\eps
\nonumber\\
   &\quad\times
	\bigg\{-\frac{1}{\eps}  + 
	2 (1 + z^2) \ln z_+
	- (2 + z^2) \ln\Big(\frac{1 + z^2}{16}\Big) - 
     	2 \sqrt{1 + z^2} + 1\bigg\}\,,
\end{align}
with $z_+ = (\sqrt{1+z^2}+1)/4$ and $\bar\tau=  \tau e^{\gamma_E}$.

\subsubsection*{Wilson-line diagram}

The calculation of the second diagram in Figure \ref{jet:diagrams} proceeds along the same lines, but this time it will be essential to keep the analytic regulator non-zero. We now start from
\begin{align}
   {\cal J}_{L}^{(1b)}(b,p^\perp) 
   &= \alpha_s C_F\,\frac{2^{1+2\eps}}{\pi^{2-2\eps}} \;\,\mutilde^{2\eps} \int d^dq \; \delta(q^2)\,\theta(q^0)\;
	\int d^dk \,\left(\frac{\nu_+}{k_+}\right)^\alpha \delta(k^2)\,\theta(k^0)
\\
   &\quad\times\,
    \frac{q_-(k_-+q_-)}{k_-\,(2 k\cdot q)}\;\,
    \delta(Q-q_--k_-)\;
    \delta^{d-2}(p^\perp-q^\perp-k^\perp)\;
    \delta\Big(b-\frac12 |q^\perp| - \frac12 |k^\perp|\Big)\,,\nonumber
\end{align}
which results in
\begin{align}\label{jet:wl:bp}
   {\cal J}_{L}^{(1b)}(b,p^\perp) 
   &= \frac{\alpha_s C_F}{2\pi^{1-\eps}}\,
   \left(\frac{\mu^2e^{\gamma_E}}{b^2}\right)^\eps \;\frac{\Omega_{d-3}}{b}
   \;\left(\frac{\nu_+Q}{b^2}\right)^\alpha (1-y^2)^{-1-\eps} \;I_L^{(1b)}(y)
\end{align}
with the two-dimensional integral
\begin{align}\label{jet:wl:inty}
   I_L^{(1b)}(y) &= \frac{\sqrt{1-y^2}}{\pi} y^{2\eps}\!\!
   	\int_0^1 \!\!d\eta \,\eta(1-\eta)^{-1+\alpha}\!\!
	\int_{1-y}^{1+y} \!\!\!\!d\xi\; 
	\frac{\xi (2-\xi)^{1-2\alpha} (1+y-\xi)^{-\frac12-\eps}(\xi-1+y)^{-\frac12-\eps}}
	{(\xi-2y\eta)^2+4\eta(1-y)(1+y-\xi)}\,.
\end{align}
Because of the singularity of the integrand for $\eta\to 1$ this integral would be ill-defined without the additional analytic regulator $\alpha$. In physical terms this is the limit where the minus component of the gluon momentum, which we assumed to be of the order of the large scale $Q$, vanishes. We then expand in the analytic regulator using the relation
\begin{equation}
  (1-\eta)^{-1+\alpha} = \frac{1}{\alpha} \,\delta(1-\eta) 
+ \left[ \frac{1}{1-\eta} \right]_+ + \calO(\alpha)\,.
\end{equation}
The subsequent calculation follows the same steps as before, but it is considerably more involved. In particular, the exact result in terms of hypergeometric functions now becomes rather lengthy. All we need here is the full expression in the limit $y\to1$, as well as the leading term in the $\eps$-expansion for arbitrary $y<1$. In the collinear limit, we now find
\begin{align}\label{jet:se:inty1}
   I_L^{(1b)}(y\to 1) &\simeq 
	- 2^{1 - 2 \eps} \;\frac{\Gamma^2(2 - \eps)}{\eps\,\Gamma(3 - 2 \eps)} 
	+ (1-y)^{-\eps}\;2^{1 - \eps}\;\frac{\Gamma(1 - 2 \eps)\, \Gamma(1 + 2 \eps)}{\Gamma(1 - \eps)\,\Gamma(1 + \eps)}
\nonumber\\
   &\quad\times
	\left\{ \frac{1}{\alpha} - \frac{1}{2\eps} - \ln(1-y) - 3 \ln2 + \Psi\Big(\frac12 + \eps\Big) - \Psi(\eps) \right\},
\end{align}
where $\Psi(x)=\Gamma'(x)/\Gamma(x)$ denotes the digamma function. In contrast to the self-energy diagram, our result for the jet function in (\ref{jet:wl:bp}) now contains non-trivial distributions of the form $(1-y)^{-1-\eps}$, $(1-y)^{-1-2\eps}$ and $(1-y)^{-1-2\eps}\ln(1-y)$, which shows that one cannot naively perform the $\eps$-expansion to extract the singular terms. There is no such complication for the terms with $y<1$, where we obtain
\begin{align}\label{jet:se:inty2}
   I_L^{(1b)}(y<1) &= 	
	\frac{1}{\alpha} \Big(2-\sqrt{1-y^2}\Big)
	+ 2\Big(2 + \sqrt{1 - y^2}\Big)  \ln \Big(\frac{1 + \sqrt{1 - y^2}}{2} \Big)
\nonumber\\
   &\quad
	+ \frac{1 - 8 y + y^2}{2 y}  \,\ln (1-y)
	- \frac{1 + 8 y + y^2}{2 y}  \,\ln (1+y)
	+ \sqrt{1-y^2} - 1 + \calO(\eps)\,.
\end{align}
We next combine the two representations in (\ref{jet:se:inty1}) and (\ref{jet:se:inty2}) to account for the non-trivial structure of the integral in distribution space. The subsequent Laplace and Fourier transformations can again be done analytically. Our final result for the Wilson-line diagram takes the form
\begin{align}\label{jet:wl:tauz}
   \overline{\cal J}_{L}^{(1b)}(\tau, z) 
   &=    \overline{\cal J}^{(0)}(\tau, z) \,
	\frac{\alpha_s C_F}{4\pi}\;\big( \mu^2 \bar \tau^2\big)^\eps\; \big(\nu_+ Q \bar \tau^2\big)^\alpha
\\
   &\quad\!\times
	\bigg\{\!\!
	- \frac{2}{\alpha} 
		\Big( \frac{1}{\eps} + 2 \ln z_+\Big)
		+ \frac{2}{\eps^2}  + \frac{2}{\eps} 
	-8\dilog\Big(\! -\frac{z_-}{z_+} \Big) 
	+ 8\dilog( -\sqrt{1+z^2})
	 - 4 \ln^2 z_+
\nonumber\\
   &\hspace{1.05cm}
	+ \ln^2(1 + z^2) + 2 z^2 \ln(1 + z^2)
	+ 4 (1 - z^2) \ln(4z_+)
	  + 4\sqrt{1 + z^2} - 8 \ln 2 - \frac{\pi^2}{6}\bigg\}\nonumber,
\end{align}
where again $z_\pm = (\sqrt{1+z^2}\pm1)/4$ and $\bar\tau=  \tau e^{\gamma_E}$.

\subsubsection*{Right jet function}

We now briefly summarize our results for the right jet function. First of all, we have seen above that the self-energy diagram does not generate a divergence in the additional regulator and, consequently, we obtain the same contribution as for the left jet function,
\begin{align}\label{jet:se:R}
\overline{\cal J}_{R}^{(1a)}(\tau, z) =\overline{\cal J}_{L}^{(1a)}(\tau, z)\,.
\end{align}
The Wilson-line diagram then follows from (\ref{jet:wl:bp}) and (\ref{jet:wl:inty}) with the replacement
\begin{equation}
\left(\frac{\nu_+(1-\eta)Q}{(2-\xi)^2b^2}\right)^\alpha \,\to\;\,
 \left(\frac{\nu_+}{(1-\eta)Q}\right)^\alpha.
\end{equation}
Note that the phase-space regulator now enters with the large scale, since $k_+\sim Q$ in the anti-collinear region. Without going into further details, the exact expression in broadening-momentum space becomes 
\begin{align}
   I_R^{(1b)}(y\to 1) &\simeq 
	- 2^{1 - 2 \eps} \;\frac{\Gamma^2(2 - \eps)}{\eps\,\Gamma(3 - 2 \eps)} 
	+ (1-y)^{-\eps}\;2^{1 - \eps}\;\frac{\Gamma(1 - 2 \eps)\, \Gamma(1 + 2 \eps)}{\Gamma(1 - \eps)\,\Gamma(1 + \eps)}
\nonumber\\
   &\quad\times
	\left\{ - \frac{1}{\alpha} - \frac{1}{2\eps} + \ln(1-y) - \ln2 -2 \gamma_E - 2\Psi(2\eps) \right\}
\end{align}
in the collinear limit. For $y<1$ we get
\begin{align}
   I_R^{(1b)}(y<1) &= 	
	- \frac{1}{\alpha} \Big(2-\sqrt{1-y^2}\Big)
	+ \frac{1 + y^2}{2 y}  \,\ln \Big(\frac{1-y}{1+y}\Big)
	+ \sqrt{1-y^2} - 1 + \calO(\eps)\,.
\end{align}
After Laplace and Fourier transformation this translates into
\begin{align}\label{jet:wl:R}
   \overline{\cal J}_{R}^{(1b)}(\tau, z) 
   &=    \overline{\cal J}^{(0)}(\tau, z) \,
	\frac{\alpha_s C_F}{4\pi}\;\big( \mu^2 \bar \tau^2\big)^\eps\; \big(\nu_+/ Q \big)^\alpha
\nonumber\\
   &\quad\times
	\bigg\{
	 \frac{2}{\alpha} 
		\Big( \frac{1}{\eps} + 2 \ln z_+ \Big)
		+ \frac{2}{\eps} + 8 \dilog( -\sqrt{1+z^2}) 
		+ \ln^2(1 + z^2) + 2 z^2 \ln(1 + z^2)
\nonumber\\
   &\hspace{1.3cm}
		+ 4 (1 - z^2) \ln(4z_+)  + 4\sqrt{1 + z^2} - 8 \ln 2 + \frac{2\pi^2}{3}
		\bigg\}\,.
\end{align}

\subsection{Remainder function $W$\label{sec:W}}

We are now ready to combine our results for the jet and soft functions. According to (\ref{JJSfact}), the product of jet and soft functions determines the anomaly exponent $F_B(\tau,z,\mu)$ and the remainder function $W(\tau_L,\tau_R,z_L,z_R,\mu)$. As the one-loop anomaly exponent has already been given in our previous work \cite{Becher:2011pf}, we focus here on the extraction of the remainder function. The renormalized function $W = Z_W W^{\rm{bare}}$ fulfils the renormalization group (RG) equation \cite{Becher:2011pf}
\begin{align}\label{RGEW}
  \frac{d}{d\ln\mu}\,W(\tau_L,\tau_R,z_L,z_R,\mu) 
   &= \Big[ 2\Gamma_{\rm cusp}^F(\alpha_s)\,\ln\big( \mu^2\bar\tau_L\bar\tau_R \big)
    - 4\gamma^q(\alpha_s) \Big]\,W(\tau_L,\tau_R,z_L,z_R,\mu) \,,
\end{align}
where $\Gamma_{\mathrm{cusp}}$ denotes the cusp anomalous dimension and $\gamma^q$ refers to the quark anomalous dimension as defined in \cite{Becher:2009qa}. The Z-factor $Z_W$ fulfils the same RG equation. To one-loop order, the solution takes the form
\begin{align}
 Z_W = 1 + \frac{\alpha _s}{4 \pi } \left[-\frac{\Gamma_0^F }{\epsilon ^2}
-\frac{1}{\epsilon }\Big(\Gamma_0^F \ln( \mu^2\bar\tau_L\bar\tau_R)-2\gamma^{q}_0\Big)\right]
+\calO(\alpha_s)^2\,,
\end{align}
where $\Gamma_0^F=4C_F$ and $\gamma^{q}_0=-3C_F$ are the leading coefficients in the perturbative expansions
$\Gamma_{\mathrm{cusp}}= \sum_{n=0}^\infty \Gamma^F_n (\frac{\alpha_s}{4\pi})^{n+1}$ and 
$\gamma^{q} = \sum_{n=0}^\infty \gamma^{q}_n \,(\frac{\alpha_s}{4\pi})^{n+1}$.  To the given accuracy, it is furthermore convenient to rewrite the remainder function as a product of a left and a right function,
\begin{align}\label{eq:wfun}
  W(\tau_L,\tau_R,z_L,z_R,\mu) &= 
	w(\tau_L,z_L,\mu)\, w(\tau_R,z_R,\mu) + \calO(\alpha_s^2)\,.
\end{align}
Combining our results for the jet and soft functions from the previous section, we find
\begin{align}
w(\tau,z,\mu) &= \overline{\cal J}^{(0)}(\tau, z) \,\bigg\{ 1 
	+ \frac{\alpha_s C_F}{4\pi}\bigg[ \ln^2( \mu^2\bar\tau^2) + 3\ln( \mu^2\bar\tau^2) 
	-8 \dilog\Big(\! -\frac{z_-}{z_+} \Big) + 16 \dilog( -\sqrt{1+z^2}) 
\nonumber\\
   &\hspace{4.5cm}
	- 4 \ln^2 z_+ + 2\ln^2(1 + z^2) + 2 (5-3z^2) \ln(4z_+) + 6\sqrt{1 + z^2}
\nonumber\\
   &\hspace{4.5cm}
	-(2-3z^2) \ln(1 + z^2)  - 12 \ln 2 +1 + \frac{\pi^2}{2}\bigg]\bigg\}\,.
\end{align}
The fact that our result for the remainder function $W$ is left-right symmetric and that it fulfils the RG equation (\ref{RGEW}) provides a non-trivial check of our calculation.

Later, we will need logarithmic moments of the remainder function. For the lowest moments, we find
\begin{align}\label{wmoments}
\int_0^\infty\! dz\, w(\tau,z,\mu) &=1 + \frac{\alpha_s C_F}{4\pi}\bigg[\ln^2( \mu^2\bar\tau^2) + 3\ln( \mu^2\bar\tau^2)+ 1-\frac{5 \pi ^2}{6} \bigg]\,, \nonumber \\
\int_0^\infty \! dz\,\ln z_+\, w(\tau,z,\mu)  &= \frac{\alpha_s C_F}{4\pi}\bigg[6+\frac{\pi ^2}{2} -20 \zeta _3+\frac{8}{3} \pi ^2 \ln2 -12  \ln^22 -\frac{32 }{3}\ln^32\bigg]\,.
\end{align}
These are used to compute the fixed-order expansion of the resummed results in Section \ref{sec:fixed}.

\section{Two-loop anomaly coefficient\label{sec:twoloop}}

\subsection{Setup of the calculation}

The anomaly coefficient $F_B(\tau,z,\mu)$ can be extracted from the divergences in the analytic regulator of any of the soft and jet functions. Here, we will consider the two-loop soft function, since in this case there are again up to two partons in the final state and the integrals are similar to the ones that we encountered in the calculation of the one-loop jet function. As the divergences cancel in each hemisphere independently, it is sufficient to consider emissions into one of the hemispheres only. To be specific, we focus on the emissions into the left hemisphere.

\begin{figure}[t!]
\begin{center}
\begin{tabular}{ccccccc}
\includegraphics[width=0.19\textwidth]{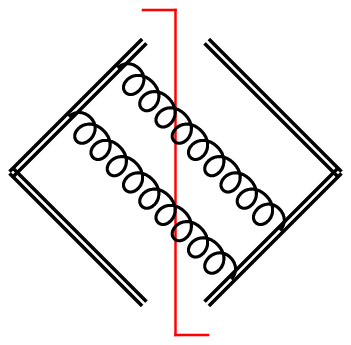} & & 
\includegraphics[width=0.19\textwidth]{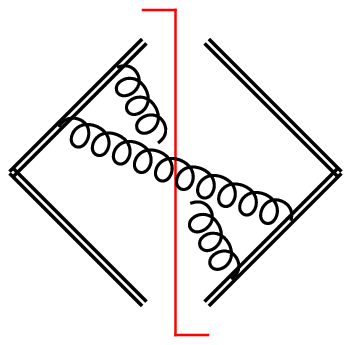} & &
\includegraphics[width=0.19\textwidth]{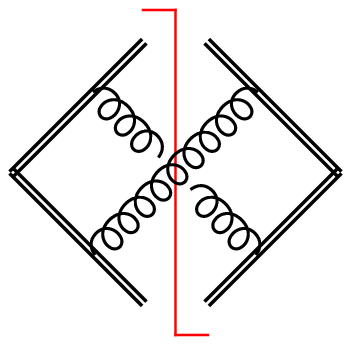} & & 
\includegraphics[width=0.19\textwidth]{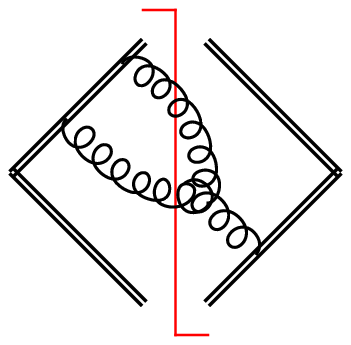}
\\[0.2em]
(a) & & 
(b) & & 
(c) & & 
(d)
\\[1.2em]
\includegraphics[width=0.19\textwidth]{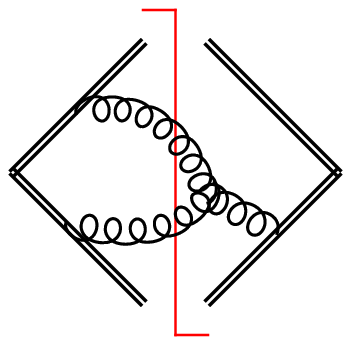} & & 
\includegraphics[width=0.19\textwidth]{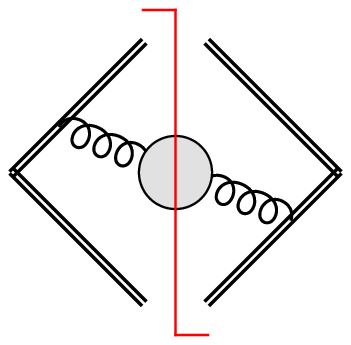} & &
\includegraphics[width=0.19\textwidth]{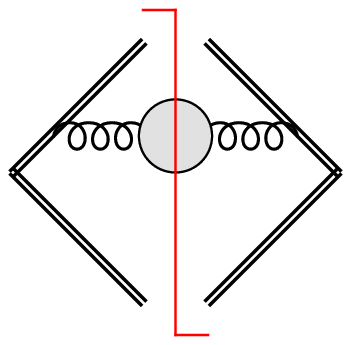} & & 
\includegraphics[width=0.19\textwidth]{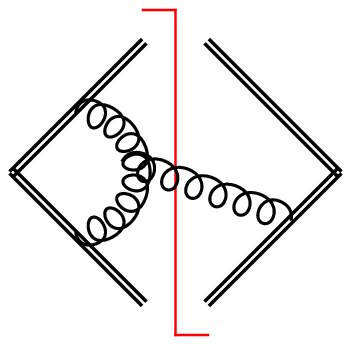}
\\[0.2em]
(e) & & 
(f) & & 
(g) & & 
(h)
\end{tabular}
\end{center}
\vspace{-0.5cm}
\caption{\label{soft:nnlodiagrams}
Diagrams that give non-vanishing contributions to the soft function at NNLO. In addition there are mirror-symmetrical graphs, which we take into account by multiplying each diagram with a symmetry factor $s_i$, where $s_a=s_b=s_f=s_g=2$, $s_c=1$ and $s_d=s_e=s_h=4$, see text.
}
\end{figure}

At two-loop order the purely virtual corrections are again scaleless and vanish. Among the mixed virtual-real and double real emissions, only the diagrams in Figure \ref{soft:nnlodiagrams} give non-vanishing contributions. The same matrix elements also arise in other two-loop computations of soft functions \cite{Belitsky:1998tc,Becher:2005pd,Kelley:2011ng,Monni:2011gb,Hornig:2011iu,Li:2011zp,Kelley:2011aa,Becher:2012za,Ferroglia:2012uy}, what makes our case different are the phase-space constraints and the necessity of working with an additional analytic regulator. In the following, we denote the individual contributions of these diagrams to the soft function by
\begin{equation}
    {\cal S}_L^{(2i)}(b_L,b_R,p_L^\perp, p_R^\perp),
	\quad\qquad i=a,b,\ldots,h
\end{equation}
and similarly for the Laplace-Fourier transformed soft function $\overline{\cal S}$.

The diagrams in Figure \ref{soft:nnlodiagrams} have counterpart diagrams that follow by exchanging $n^\mu\leftrightarrow\bar{n}^\mu$ as well as complex conjugation. It turns out that the matrix elements of diagrams (d), (e) and (g) are not symmetric under $n^\mu\leftrightarrow\bar{n}^\mu$. In an integral over a symmetric phase space, these mirror graphs would nevertheless give the same result as the original diagrams. As our regulator (\ref{regulator}) breaks this symmetry, it is however not guaranteed that the contributions from the mirror graphs and the original diagrams agree. We therefore explicitly calculated these mirror graphs and found that they do give the same results as far as the divergences in the phase-space regulator $\alpha$ are concerned. The contributions from these mirror graphs are therefore included in the symmetry factors $s_i$.

Before going over to the more technical aspects of the NNLO calculation, we give the result for the one-particle cut in the last diagram of Figure \ref{soft:nnlodiagrams}. As there is only one soft particle in the final state, this diagram can easily be calculated along the lines of the NLO calculation. In broadening-momentum space, we find
\begin{align}
   {\cal S}_L^{(2h)}(b_L,b_R,p_L^\perp, p_R^\perp) 
   &= \frac{\alpha_s^2}{4\pi^{3-\epsilon}} \,C_A C_F \left( \mu^2 e^{\gamma_E} \right)^{2\epsilon}
	\frac{\nu_+^\alpha}{\alpha}\,
	\frac{\Gamma^2(1+\eps)\Gamma^3(-\eps)}{\Gamma(-2\eps)}\,\cos(\pi\eps)\,
    \nonumber\\
	&\quad\times
	 p_L^{-2-2\eps-\alpha}\,
    \delta\Big(b_L-\frac{p_L}{2}\Big)\,\delta(b_R)\,\delta^{d-2}(p_R^\perp)
    \,,
\end{align}
which after Laplace and Fourier transformation turns into
\begin{align}\label{eq:2h}
   &\overline{\cal S}_L^{(2h)}(\tau_L,\tau_R,z_L, z_R) 
   = \left(\frac{\alpha_s}{4\pi}\right)^2 C_A C_F
	\big( \mu^2 \bar \tau_L^2\big)^{2\eps}
	\frac{\big(\nu_+ \bar \tau_L\big)^\alpha}{\alpha}
	\nonumber\\
   &\quad\times\bigg\{\!
	- \frac{4}{\eps^3}  -\frac{16\ln z_+}{\eps^2}
	- \bigg( 48\dilog\Big(\! -\frac{z_-}{z_+} \Big) +32\ln^2 z_+ +\frac{10\pi^2}{3} \bigg)\frac{1}{\eps}
	+288 \nielsen\Big(\! -\frac{z_-}{z_+} \Big) 
	\nonumber\\
   &\hspace{1.2cm}
	-48 \trilog\Big(\! -\frac{z_-}{z_+} \Big) 
	-\frac{128}{3} \ln^3 z_+ 
	- 192 \ln z_+ \,\dilog\Big(\! -\frac{z_-}{z_+} \Big)
	-\frac{40\pi^2}{3} \ln z_+ - \frac{224}{3} \zeta_3 \bigg\}\,
\end{align}
 up to finite terms in the $\alpha$-expansion that are not needed to extract the anomaly coefficient. The function $\nielsen(z)$ is a Nielsen generalized polylogarithm. Since we only consider emissions into the left hemisphere, the result only depends on $z_L$ and we therefore drop the superscript $L$ on $z_\pm^L= (\sqrt{1+z_L^2}\pm1)/4$ throughout this section for brevity. Next, we will discuss the calculation of the two-particle cut diagrams in detail. We will then explain how to extract the bare anomaly coefficient from the two-loop soft function, discuss its renormalization and give our final two-loop result.

\subsection{Two-particle cut diagrams\label{sec:twoparticlecut}}

For the two-particle cut diagrams, we only need to consider the case where both of the gluons are emitted into the left hemisphere. Similar to the one-loop jet function, the calculation is complicated by a non-trivial angle in the transverse-momentum plane. We organize our calculation along the same lines as in Section \ref{sec:jet}. It turns out, however, that one has to compute the two-loop soft integrals to one order higher in the $\eps$-expansion, which severely complicates the calculation.

As an example of a two-particle cut diagram, we will consider diagram (b) in detail. The calculation of the remaining diagrams in Figure \ref{soft:nnlodiagrams} proceeds similarly, and the results are summarized in Appendix \ref{sec:diagrams}. As in the calculation of the one-loop jet function, we checked our analytic results with a purely numerical approach. The details of the numerical calculation can be found in Appendix \ref{sec:numerical:soft}.

We now turn to the evaluation of diagram (b). In broadening-momentum space we start from
\begin{align}
&{\cal S}_L^{(2b)}(b_L,b_R,p_L^\perp, p_R^\perp)
   = \alpha_s^2 C_F\Big(C_F-\frac{C_A}{2}\Big)\;\mutilde^{4\eps}\!\!
	\int d^dk \,\left(\frac{\nu_+}{k_+}\right)^\alpha \delta(k^2)\,\theta(k^0)\,
	\int d^dl \,\left(\frac{\nu_+}{l_+}\right)^\alpha \delta(l^2)\,\theta(l^0)
\nonumber\\
   &\quad\times
    \frac{2^{1+4\eps}}{\pi^{4-4\eps}} \;
    \frac{\theta(k_--k_+)\theta(l_--l_+)}{k_+(k_++l_+)k_-(k_-+l_-)}\;
    \delta^{d-2}(p_L^\perp-k^\perp-l^\perp)\,
    \delta\Big(b_L-\frac12 |k^\perp| - \frac12 |l^\perp|\Big)\,
    \delta^{d-2}(p_R^\perp)\,
    \delta(b_R).
\end{align}
We now use the first delta function in the second line to perform the $l^\perp$-integration and the on-shell conditions for the integrations over the minus components. We then arrive at
\begin{align}
	&{\cal S}_L^{(2b)}(b_L,b_R,p_L^\perp, p_R^\perp)
	=  \alpha_s^2 C_F\Big(C_F-\frac{C_A}{2}\Big)\,
		\frac{2^{-1+4\eps}}{\pi^{4-4\eps}}\;\mutilde^{4\eps}
		\int_0^\infty \!dk\; k^{d-5}\;\int_0^k \!dk_+ \; 
		\int_0^{l_+^{\rm max}} \!\!\!\!dl_+ \; 
\nonumber\\
   &\quad\times
   \int_{-1}^1 \! d\cos\theta\; \sin^{d-5}\theta\;
   \int d\Omega_{d-3}\,
    \left(\frac{\nu_+^2}{k_+l_+}\right)^\alpha
    \frac{1}{k^2 l_++(p_L^2 + k^2 - 2 p_L k \cos \theta)k_+}
\nonumber\\
   &\quad\times
    \frac{1}{k_++l_+}\;
    \delta\Big(b_L-\frac{1}{2}\,k - \frac12 \sqrt{p_L^2+k^2-2p_L k \cos\theta}\Big)\,
    \delta^{d-2}(p_R^\perp)\;
    \delta(b_R)\,,
\end{align}
with $l_+^{\rm max}=\sqrt{p_L^2+k^2-2p_L k \cos\theta}$, where $\theta$ denotes the angle between $\vec{k}^\perp$ and $\vec{p}_L^\perp$, and we write $k=|k^\perp|$ and $p_L=|p_L^\perp|$.  Next, we use the remaining delta function to perform the integration over $\cos\theta$. Substituting $l_+=\rho k_+$, one may easily perform the $k_+$-integration. We furthermore introduce dimensionless variables via $k = b_L \xi$ and $y = p_L/2 b_L$, which results in
\begin{align}\label{soft:b:bp1}
	&{\cal S}_L^{(2b)}(b_L,b_R,p_L^\perp, p_R^\perp)
\nonumber\\
   &\quad
= -\frac{\alpha_s^2 C_F}{2\pi^{3-2\eps}}\, \Big(C_F-\frac{C_A}{2}\Big)\,
   \left(\frac{\mu^2e^{\gamma_E}}{b_L}\right)^{2\eps} \frac{\Omega_{d-3}}{b_L^3}\;\frac{1}{\alpha}\,
   \left(\frac{\nu_+}{b_L}\right)^{2\alpha} \!(1-y^2)^{-1-\eps} \,
    \delta^{d-2}(p_R^\perp)\;
    \delta(b_R)\,I_S^{(2b)}(y)\,,
\end{align}
with the two-dimensional integral
\begin{align}\label{soft:b:bp2}
   I_S^{(2b)}(y) &= \frac{\sqrt{1-y^2}}{\pi} \;y^{2\eps}\!
    \int_{1-y}^{1+y} \!\!d\xi\;\xi^{-1-2\alpha}\, (2-\xi)\! 
    \int_0^{\frac{2-\xi}{\xi}} d\rho \;\frac{\rho^{-\alpha}}{1+\rho} \,
	\frac{ (1+y-\xi)^{-\frac12-\eps}(\xi-1+y)^{-\frac12-\eps}}
	{(2-\xi)^2+\rho\xi^2}\,.
\end{align}
Note that we again factored out a singular distribution in (\ref{soft:b:bp1}), and that the remaining integral is finite in the limit $\alpha\to0$. The subsequent calculation proceeds along the same lines as in Section \ref{sec:jet}. In particular, setting $\alpha=0$ we could again find an exact representation of the above integral in terms of hypergeometric functions. Again, we need the exact expression only in the limit $y\to1$. We find
\begin{align}
   I_S^{(2b)}(y\to 1) &\simeq 
	 -(1-y)^{-\eps}\;2^{-1 - \eps}\;\frac{\Gamma(1 - 2 \eps)\, \Gamma(1 + 2 \eps)}{\Gamma(1 - \eps)\,\Gamma(1 + \eps)}
	\left\{   \ln(1-y) - \ln2 - \Psi\Big(\frac12 + \eps\Big) -\gamma_E \right\}.
\end{align}
In addition, we need the first two coefficients in the $\eps$-expansion for arbitrary $y<1$. To this end, we use the Mathematica package HypExp \cite{Huber:2005yg,Huber:2007dx}, which allows one to expand hypergeometric functions around integer and half-integer parameters. The extension to half-integer parameters is currently restricted to special classes of hypergeometric functions. As some of the functions that enter our calculation do not belong to these classes (these are of the type $3^2_1$ in the notation of \cite{Huber:2007dx}),  we use an integral representation for these hypergeometric functions to perform the $\eps$-expansion manually. We could moreover make use of results from \cite{Davydychev:2003mv} for some special cases such as ${}_3F_2(1,\frac32,\frac32;2+\eps,\frac32+\eps;y^2)$. The expansion of the above integral for arbitrary $y<1$ finally reads
\begin{align}
   I_S^{(2b)}(y<1) &= 	
	\frac{\sqrt{1-y^2}}{2y} \arcsin y - \frac12 \ln(1-y^2)
	+ \bigg\{ \dilog \Big( \frac{1-\sqrt{1-y^2}}{1+\sqrt{1-y^2}} \Big)
	- \dilog \Big( \frac{\sqrt{1-y^2}-1}{1+\sqrt{1-y^2}} \Big)
\nonumber\\
   &\quad
	+ \frac{\sqrt{1-y^2}}{2y} \Big( 4 {\rm Cl}_2\big(\pi-\arcsin y\big)- {\rm Cl}_2\big(\pi-2\arcsin y\big) \Big)
\nonumber\\
   &\quad
	-\ln(1-y^2) \ln(1+\sqrt{1-y^2}) 
	+ \frac12 \ln^2(1-y^2) \bigg\} \eps 
	+ \calO(\eps^2)\,,
\end{align}
where ${\rm Cl}_2(x)$ denotes the Clausen function
\begin{align}
 {\rm Cl}_2(x)  = - \int_0^x dt \;\ln \big( 2\sin(t/2)\big) 
		= \frac{i}{2} \Big( \dilog\big(e^{-ix}\big) - \dilog\big(e^{ix}\big) \Big)\,.
\end{align}
After Laplace and Fourier transformation this translates into
\begin{align}
&\overline{\cal S}_L^{(2b)}(\tau_L,\tau_R,z_L, z_R) 
   =   
\left(\frac{\alpha_s}{4\pi}\right)^2 C_F\Big(C_F-\frac{C_A}{2}\Big)
	\big( \mu^2 \bar \tau_L^2\big)^{2\eps}
	\frac{\big(\nu_+ \bar \tau_L\big)^{2\alpha}}{\alpha}
	\nonumber\\
   &\quad\times\bigg\{\!
	\frac{2}{\eps^3}  +\frac{8\ln z_+}{\eps^2}
	+ \bigg( 24\dilog\Big(\! -\frac{z_-}{z_+} \Big) +16\ln^2 z_+ +3 \pi^2 \bigg)\frac{1}{\eps}
	+64\,h_3(z_L)-80 \nielsen\Big(\! -\frac{z_-}{z_+} \Big) 
	\nonumber\\
   &\hspace{1.2cm}
	+56 \trilog\Big(\! -\frac{z_-}{z_+} \Big)  
	+96 \ln z_+ \,\dilog\Big(\! -\frac{z_-}{z_+} \Big)
	+\frac{64}{3} \ln^3 z_+
	+4\pi^2 \ln \Big(\frac{z_+}{4}\Big) + \frac{112}{3} \zeta_3 \bigg\},
\end{align}
which again holds up to finite terms in the $\alpha$-expansion. The definition of the function $h_3(z)$ can be found in equation (\ref{eq:def:h3}) below. It involves elliptic integrals and cannot be expressed in terms of polylogarithmic functions.

\subsection{Extraction of the anomaly coefficient\label{sec:anomaly}}

In our previous work \cite{Becher:2011pf}, we showed that, to all orders in perturbation theory, the logarithm of the soft function can be written in the form
\begin{align}\label{eq:logS:FB}
 \ln \overline{\cal S}(\tau_L,\tau_R,z_L, z_R) 
 	= 2F_B(\tau_L,z_L) \ln (\nu_+ \bar \tau_L) - 2F_B(\tau_R,z_R) \ln (\nu_+ \bar \tau_R) 
	+ k_0^S(\tau_L,\tau_R,z_L, z_R) \,,
\end{align}
where we suppressed the divergent terms in the analytic regulator that cancel in the product of the jet and soft functions. The function $k_0^S$ contributes to the remainder function $W$ and is therefore irrelevant for our purposes. In our two-loop calculation, we restricted our attention to emissions into the left hemisphere and we focused on the divergent terms in the analytic regulator. We will now show that this is sufficient to extract the anomaly coefficient $F_B(\tau,z)$. 

Making the divergences in the analytic regulator explicit, the logarithm of the soft function can be rewritten in the form
\begin{align}\label{eq:logS}
 \ln \overline{\cal S}(\tau_L,\tau_R,z_L, z_R) 
 	= \sum_{n=1}^{\infty} \;(\nu_+\bar \tau_L)^{n\alpha} \; 
	\Big( \frac{1}{\alpha} \,f_n(\tau_L,z_L) + r_n(\tau_L,z_L) \Big) +\ldots\,,
\end{align}
where the ellipsis denote terms that involve the variables of the right hemisphere. According to our phase-space regulator (\ref{regulator}), an $n$-particle cut diagram involves a power $(\nu_+)^{n\alpha}$. The functions $f_n$ and $r_n$ are therefore of $\calO(\alpha_s^n)$. Expanding the above decomposition to two-loop order and keeping only terms that are relevant for the extraction of the anomaly coefficient, we arrive at
\begin{align}\label{eq:logSb}
 \ln \overline{\cal S}(\tau_L,\tau_R,z_L, z_R) 
 	= \Big(f_1(\tau_L,z_L) + f_2(\tau_L,z_L)\Big)\frac{1}{\alpha}
	+\Big(f_1(\tau_L,z_L) + 2f_2(\tau_L,z_L)\Big)\ln (\nu_+ \bar \tau_L) 
	+\ldots\,.
\end{align}
Notice that the coefficients of the divergence and the logarithm do not agree beyond one-loop order. This is different from the more common situation in dimensional regularization, where the renormalization scale is tied to the coupling constant in the form $\alpha_s \mu^{2\eps}$. In contrast, the two-loop diagrams in our calculation do not have a homogeneous scaling in $\nu_+$, since they involve both one-particle and two-particle cuts. 

From (\ref{eq:logS:FB}) we see that the anomaly coefficient is determined by the logarithm rather than the divergence. Up to two-loop order it is given by the combination
\begin{align}
F_B(\tau,z)= \frac12 f_1(\tau,z) + f_2(\tau,z).
\end{align}
Taking the exponent of (\ref{eq:logS}) and expanding to second order in $\alpha_s$, we may then extract the bare anomaly coefficient from the single logarithmic terms in $\nu_+$. As this also generates a crossed term between the one-loop terms in $f_1$ and $r_1$, we first need to determine their higher-order terms in $\eps$. We find
\begin{align}
   f_1(\tau,z) 
   &= \frac{\alpha_s C_F}{\pi}
	\big( \mu^2 \bar \tau^2\big)^{\eps}\;
	\bigg\{
	\frac{1}{\eps}  + 2\ln z_+
	+ \bigg( 2\dilog\Big(\! -\frac{z_-}{z_+} \Big) +2\ln^2 z_+ +\frac{\pi^2}{4} \bigg)\eps
	+ \bigg( 2 \trilog\Big(\! -\frac{z_-}{z_+} \Big)   
	\nonumber\\
   &\quad
	-4 \nielsen\Big(\! -\frac{z_-}{z_+} \Big)
	+\frac{4}{3} \ln^3 z_+ 
	+ 4 \ln z_+ \,\dilog\Big(\! -\frac{z_-}{z_+} \Big)
	+\frac{\pi^2}{2} \ln z_+ + \frac{7}{3} \zeta_3 \bigg) \eps^2 
	+\calO(\eps^3)\bigg\}
\end{align}
and
\begin{align}
   r_1(\tau,z) 
   &= \frac{\alpha_s C_F}{4\pi}
	\big( \mu^2 \bar \tau^2\big)^{\eps}\;
	\bigg\{\!
	-\frac{2}{\eps^2}  
	+  8\dilog\Big(\! -\frac{z_-}{z_+} \Big) +4\ln^2 z_+ +\frac{5\pi^2}{6} 
	+ \bigg( 8 \trilog\Big(\! -\frac{z_-}{z_+} \Big)   
	\\
   &\quad
	-32 \nielsen\Big(\! -\frac{z_-}{z_+} \Big)
	+\frac{16}{3} \ln^3 z_+ 
	+ 24 \ln z_+ \,\dilog\Big(\! -\frac{z_-}{z_+} \Big)
	+\frac{8\pi^2}{3} \ln z_+ + \frac{34}{3} \zeta_3 \bigg) \eps 
	+\calO(\eps^2)\bigg\}.
	\nonumber
\end{align}
We next renormalize the coupling constant in the $\overline{\rm MS}$ scheme, and recall that the renormalized anomaly coefficient fulfils the RG equation
\cite{Becher:2011pf}
\begin{align}
  \frac{d}{d\ln\mu}\,F_B(\tau,z,\mu) 
   &= \Gamma_{\rm cusp}(\alpha_s) \,.
\end{align}
The anomaly coefficient renormalizes additively, $F_B = F_B^{\rm{bare}} + Z_{F_B}$, and the Z-factor fulfils the same RG equation. Up to two-loop order, the solution takes the form
\begin{align}
 Z_{F_B} = -\frac{\alpha _s}{4 \pi } \frac{\Gamma_0^F }{2\epsilon}
- \left(\frac{\alpha _s}{4 \pi }\right)^2 \left[\frac{\Gamma_1^F }{4\epsilon} - \frac{\beta_0\Gamma_0^F }{4\epsilon ^2}
\right]
+\calO(\alpha_s^3)\,.
\end{align}
The relevant anomalous dimensions are
\begin{align}
 \Gamma_0^F& =4C_F\,,  &
\frac{\Gamma_1^F}{\Gamma_0^F} 
   &= \left( \frac{67}{9} - \frac{\pi^2}{3} \right) C_A - \frac{20}{9}\,T_F n_f \,, &
\beta_0 &= \frac{11}{3}\,C_A - \frac43\,T_F n_f \,.
\end{align}
The renormalized anomaly coefficient finally takes the form
\begin{equation}
F_B(\tau,z,\mu) = \frac{\alpha_s}{4\pi}\; \Gamma_0^F \left\{ \ln(z_+ \mu\bar\tau)  +
\frac{\alpha_s}{4\pi} \left[ \beta_0 \ln^2(\mu\bar\tau)+ \left(\frac{\Gamma_1^F}{\Gamma_0^F}  + 2\beta_0 \ln z_+ \right) \ln(\mu\bar\tau)+ d^B_2(z) \right] \right\} \,,
\end{equation}
with $z_+=(\sqrt{1+z^2}+1)/4$. The nonlogarithmic piece of the anomaly exponent is encoded in the function $d^B_2(z)$, for which we find
\begin{align}\label{dB}
d_2^B(z) &= C_A \bigg\{\!
	-\frac{1+z^2}{9} \,h_1(z)
	+\frac{67+2 z^2}{9} \,h_2(z)
	-8\,h_3(z)
	+32 \nielsen\Big(\! -\frac{z_-}{z_+} \Big) 
	-8 \trilog\Big(\! -\frac{z_-}{z_+} \Big)
\nonumber\\
&\hspace{1.35cm}	
	+8 \nielsen(-w) 
	-24 \trilog(-w)
   	-24 \nielsen(1-w) 
	+8 \trilog(1-w)
   	+24 \nielsen\Big( \frac{1-w}{2} \Big) 
\nonumber\\
&\hspace{1.35cm}   	
	-8 \trilog\Big( \frac{1-w}{2} \Big)
	-8 \Big( 3 \ln z_+ + 4\ln 2\Big) \,\dilog\Big(\! -\frac{z_-}{z_+} \Big)
   	+8\ln\Big((1+w)w^3\Big) \,\dilog(-w)
\nonumber\\
&\hspace{1.35cm}      	
	-8 \ln2 \,\dilog \Big( \frac{1-w}{2} \Big)
	+4\ln\frac{w}{2}\,\ln^2 z_+ 
	+ 12 \ln^2 w \,\ln(4z_+) 
	- \frac{16}{3} \ln^3(2z_+)
\nonumber\\
&\hspace{1.35cm}     
	+ \frac{11}{3} \ln^2 z_+
	+ 16\ln2\,\ln\frac{w}{4}\,\ln z_+
	+ \Big( 24\ln^2 2 + \frac{67}{9} \Big) \ln z_+
	+ 4 \ln^2 2 \,\ln \frac{w^4}{2}
\nonumber\\
&\hspace{1.35cm}     
	+ \pi^2 \ln2
	+ \frac{290}{27}
	- 18 \zeta_3
	- \frac29 z^2
	- \frac{2w(32-z^2)}{9} \ln \Big( \frac{1+w}{w} \Big)
	- \frac{w(65 + 2z^2)}{9} \bigg\}
\nonumber\\
&\quad
+ T_F n_f \bigg\{
	\frac{2(1+z^2)}{9} \,h_1(z)
	-\frac{2(13+2 z^2)}{9} \,h_2(z)
	- \frac43 \ln^2 z_+
	- \frac{20}{9} \ln z_+
	+ \frac49 z^2
	- \frac{82}{27}
\nonumber\\
&\hspace{2.05cm}     
	+ \frac{4w(5-z^2)}{9} \ln \Big( \frac{1+w}{w} \Big)
	+ \frac{2w(11 + 2z^2)}{9} \bigg\}\,,
\end{align}
where $w = \sqrt{1+z^2}$ and $z_\pm = (w\pm1 )/4$. The result involves the three functions
\begin{align}\label{eq:def:h3}
h_1(z) &= \int_0^1 dt \;\,\frac{\arcsin t}{\sqrt{1-t^2}} \frac1{\sqrt{1+t^2 z^2}} \,,\nonumber\\
h_2(z) &= \int_0^1 dt \;\,\frac{\arcsin t}{\sqrt{1-t^2}} \sqrt{1+t^2 z^2}\,, \\
 h_3(z) &= \int_0^1 dt \;\,\frac{\arcsin t}{\sqrt{1-t^2}} \;\,\ln\big(1+\sqrt{1+t^2 z^2}\big)\,, \nonumber
\end{align}
which cannot be expressed in terms of polylogarithmic functions. They are related by the differential equations
\begin{align}
z\, h_2'(z) &= h_2(z) -h_1(z) \,,  &
z\, h_3'(z) &= \frac{\pi^2}{8} -h_1(z) \,,
\end{align}
and can be written in terms of elliptic integrals,
 \begin{align}\label{eq:elliptic}
 h_1(z) &= \frac{\pi}{2}\,  F(\mbox{$\frac{\pi}{2}$}, -z^2) 
 - \int_0^{\frac{\pi}{2}} d\theta \;F(\theta, -z^2)\,,\nonumber\\
 h_2(z) &= \frac{\pi}{2}\,  E(\mbox{$\frac{\pi}{2}$}, -z^2) 
 - \int_0^{\frac{\pi}{2}} d\theta \;E(\theta, -z^2) \,,
 \end{align}
where 
 \begin{align}
 F(\theta, x) &=\int_0^\theta dt  \;\frac1{\sqrt{1- x \sin^2 t}} \,, &
E(\theta, x) &=\int_0^\theta dt  \;\sqrt{1- x \sin^2 t} \,
\end{align}
denote the incomplete elliptic integrals of the first and second kind, respectively. For $\theta=\pi/2$, the functions are also called complete elliptic integrals and denoted by $K(x) =F(\mbox{$\frac{\pi}{2}$}, x)  $ and $E(x) = E(\mbox{$\frac{\pi}{2}$}, x)$. The integrals involving the elliptic integrals in (\ref{eq:elliptic}) can be expressed in terms of the Kamp\'e de F\'eriet hypergeometric function \cite{elliptic}.

For the numerical implementation of the resummation, the expansion of the elliptic functions around $z=\infty$ is useful. We find
\begin{align}
h_1(z) &= \frac{2 G}{z}-\frac{1}{z^2}+\frac{1-2 G}{4 z^3}+\frac{4}{9 z^4}+\dots
\,,\nonumber\\
h_2(z) &= z+\frac{G}{z}-\frac{1}{3 z^2} +\frac{1-2 G}{16 z^3}+\frac{4}{45 z^4} +\dots \,,\\
h_3(z) & =\frac{\pi ^2}{8} \ln z+\frac{7}{16} \zeta_3-\frac{\pi ^2}{8} \ln2 
   +\frac{2G}{z}-\frac{1}{2 z^2}+\frac{1-2 G}{12 z^3}
   +\frac{1}{9z^4}+ \dots\,.\nonumber
\end{align}
The expansions involve Catalan's constant $G\approx 0.915966$. 

Finally, for the fixed-order expansion of our result, we will need the integral of the product of the tree-level jet function with the anomaly exponent. We find
\begin{multline}\label{dBmoment}
\int_0^\infty dz\, \overline{\cal J}^{(0)}(z)\, d_B(z)
 =  T_F n_F\,  \left\{-\frac{14 G}{3}-\frac{34}{27}+\frac{8 \ln^2 2 }{3}+\frac{20 \ln 2 }{9}\right\} \\
+ C_A \left\{\frac{43 G}{3}-\frac{17 \zeta_3}{2}+\frac{104}{27}-\frac{16}{3} \ln^3 2-\frac{22 \ln^2 2 }{3}-\frac{64}{9} \ln 2+\pi ^2 \ln 2  \right\}  \,.
\end{multline}

\section{Resummation\label{sec:resummation}}

Our result for the cross section in Laplace space has the form
\begin{equation}\label{factlaplace}
\begin{aligned}
   \frac{1}{\sigma_0}\,\frac{d^2\sigma}{d\tau_L\,d\tau_R} 
   & = H(Q^2,\mu) \int_0^\infty\!dz_L \int_0^\infty\!dz_R\,
    \big(Q^2\bar\tau_L^2\big)^{-F_B(L_L,z_L,\mu)}\,
    \big(Q^2\bar\tau_R^2\big)^{-F_B(L_R,z_R,\mu)} 
    \\&\hspace{6cm} \times 
    W(L_L,L_R,z_L,z_R,\mu) \,.
\end{aligned}
\end{equation}
For later convenience, and with a slight abuse of notation, we have written the anomaly exponent $F_B$ as well as the remainder $W$ as functions of the logarithms
\begin{align}
L_L &= \ln(\mu \bar \tau_L) & &\text{ and} &  L_R &= \ln( \mu \bar \tau_R) \,
\end{align}
instead of $\tau_L$ and $\tau_R$. To get the resummed result, we need to solve the RG equation for the hard function and perform the Mellin inversion of the above result. 

To perform the inversion, we first rewrite the anomaly function in the form
\begin{equation}
F_B(L,z,\mu) = \frac{\alpha_s}{4\pi} \Gamma_0^F \left[ L + \ln z_+ + f_B(L,z,\mu)\right] \,,
\end{equation}
where $z_+=(\sqrt{1+z^2}+1)/4$ and 
the function
\begin{equation}
f_B(L,z,\mu) = \frac{\alpha_s}{4\pi} \left[ \beta_0 L^2 + \left(\frac{\Gamma_1^F}{\Gamma_0^F}  + 2\beta_0 \ln z_+ \right) L+ d^B_2(z) \right] + {\mathcal O}(\alpha_s^2)
\end{equation}
collects the higher-order contributions to the exponent. The anomaly can then be written in the form
\begin{equation}\label{anomalyalt}
 \big(Q^2\bar\tau^2\big)^{-F_B(L,z,\mu)} = \big(\mu \bar \tau\big)^{-\eta}  \big(z_+\big)^{- \eta_F} e^{ - 2 L \,F_B(L,z) - \eta_F f_B(L,z)}
\end{equation}
 with
 \begin{equation}
 \eta =  \eta_F = \frac{\alpha_s}{4\pi} \Gamma_0^F \ln \frac{Q^2}{\mu^2}\,.
 \end{equation}
Distinguishing $\eta$ and $ \eta_F$ in (\ref{anomalyalt}) will allow us to rewrite powers of the logarithm $L$ as derivatives with respect to $\eta$. Since the coupling constant is multiplied by a large logarithm for $\mu \sim 1/\tau$, the quantities  $\eta$ and $\eta_F$ count as ${\mathcal O}(1)$ and must be kept in the exponent. The remaining exponential on the right-hand side of (\ref{anomalyalt}) can be expanded in $\alpha_s$, since it does not contain large logarithms. After this expansion, the $\tau$-dependence of the cross section becomes very simple: at any fixed order in $\alpha_s$, it is given by a polynomial in the logarithm $L$ multiplying the factor $(\mu \bar \tau)^{-\eta}$. The Mellin inversion can then be evaluated in closed form, using the formula
 \begin{equation}
 \frac{1}{2\pi i} \int_{-i \infty + c }^{+i \infty + c }\!\!\!\!\! d\tau\, e^{b \tau} \, L^n \big(\mu \bar \tau\big)^{-\eta} = (-\partial_\eta)^n \frac{1}{b} \left(\frac{b}{\mu}\right)^\eta \frac{e^{-\gamma_E \eta}}{\Gamma(\eta)}\,.
 \end{equation}
 
Performing the inversion for the cross section, we obtain the following result for the double differential cross section:
\begin{equation}\label{factlaplacetwo}
\begin{aligned}
   \frac{1}{\sigma_0}\,\frac{d^2\sigma}{db_L\,db_R} 
   & = H(Q^2,\mu) \int_0^\infty\!dz_L \int_0^\infty\!dz_R\,  
   \big(z^L_+\big)^{- \eta_F} \big(z^R_+\big)^{-\eta_F} \, W(-\partial_{\eta_L}, -\partial_{\eta_R},z_L,z_R,\mu)\\
 &\hspace{3cm} \times \exp\left[2{ \partial_{\eta_L} \,F_B(-\partial_{\eta_L},z_L,\mu) - \eta_F f_B(-\partial_{\eta_L},z_L,\mu)} + (L\leftrightarrow R) \right] \\
 &\hspace{4cm} \times \frac{1}{b_L} \left(\frac{b_L}{\mu}\right)^{\eta_L} \frac{e^{-\gamma_E \eta_L}}{\Gamma(\eta_L)}
\, \frac{1}{b_R} \left(\frac{b_R}{\mu}\right)^{\eta_R} \frac{e^{-\gamma_E \eta_R}}{\Gamma(\eta_R)}\Bigg |_{\eta_L = \eta_R = \eta_F}\,.
 \end{aligned}
\end{equation}
Since the exponential in the second line can be expanded in $\alpha_s$ only a finite number of derivatives need to be computed, when evaluating the cross section at any given accuracy. At NNLL accuracy only the first and second derivative will be needed.

To NNLL accuracy, the general expression can be simplified further by noting that the remainder function can be written as a product of a left and a right function
\begin{equation}
W(L_L,L_R,z_L,z_R,\mu) = w (L_L, z_L,\mu)\, w(L_R, z_R,\mu) + {\mathcal O}(\alpha_s^2)\,, 
\end{equation} 
see (\ref{eq:wfun}).
To obtain compact expressions one can then define the integral
\begin{equation}
{\cal I}(-L,\mu) = \int_0^\infty\!dz_L  \big(z_+\big)^{-\eta_F}\, \exp\left[ - 2L \,F_B(L,z,\mu) - \eta_F f_B(L,z,\mu)  \right] w(L, z,\mu)\,,
\end{equation}
in terms of which the double differential cross section reads
\begin{align}
  \frac{1}{\sigma_0}\,\frac{d^2\sigma}{db_L\,db_R}  &=H(Q^2,\mu)\, {\cal I}(\partial_{\eta_L},\mu)\, {\cal I}(\partial_{\eta_R},\mu)\, \frac{1}{b_L} \left(\frac{b_L}{\mu}\right)^{\eta_L} \frac{e^{-\gamma_E \eta_L}}{\Gamma(\eta_L)}
\, \frac{1}{b_R} \left(\frac{b_R}{\mu}\right)^{\eta_R} \frac{e^{-\gamma_E \eta_R}}{\Gamma(\eta_R)}    \Bigg |_{\eta_L = \eta_R =  \eta_F}\,.
\end{align}
For the total broadening $b_T=b_L+b_R$ and the wide broadening $b_W={\rm max}(b_L,b_R)$, the expression simplifies further and we obtain
\begin{align}\label{finalres}
 \frac{1}{\sigma_0}\,\frac{d\sigma}{db_T} &= H(Q^2,\mu)\,{\cal I}^2(\partial_{\eta_T},\mu)\, \frac{e^{-\gamma_E \eta_T}}{\Gamma(\eta_T)} \,  \frac{1}{b_T} \left(\frac{b_T}{\mu}\right)^{\eta_T} \Bigg |_{\eta_T =  2\eta_F}\,,\\
 \frac{1}{\sigma_0}\,\frac{d\sigma}{db_W} &= H(Q^2,\mu)\, {\cal I}(\partial_{\eta_L},\mu)\, {\cal I}(\partial_{\eta_R},\mu)\,\frac{(\eta_L+\eta_R)\,e^{-\gamma_E (\eta_L+\eta_R)}}{\Gamma(1+\eta_L)\Gamma(1+\eta_R)}\, \frac{1}{b_W} \left(\frac{b_W}{\mu}\right)^{\eta_L+\eta_R}  \Bigg |_{\eta_L = \eta_R =  \eta_F}\,.\nonumber
 \end{align}

With the result for the cross section at hand, all that is left is to resum the logarithms in the hard function. This step is by now standard, since the same hard function appears in many processes, in particular in all dijet $e^+e^-$ event shapes and in Drell-Yan production at hadron colliders. The hard function is given by the square of the on-shell quark form factor,  $H(Q^2,\mu)=|C_V(-Q^2-i \epsilon,\mu)|^2$, which is known to three loops \cite{Baikov:2009bg,Gehrmann:2010tu}. In the following, we will suppress the $i\epsilon$-prescription attached to $Q^2$. At one-loop order, the form factor is given by
\begin{equation}
   C_V(-Q^2,\mu_h) = 1 + \frac{C_F\alpha_s(\mu_h)}{4\pi}
    \left( - L^2 + 3L - 8 + \frac{\pi^2}{6} \right) + \dots \,,
\end{equation}
where $L=\ln(-Q^2/\mu_h^2)$.
The solution of its RG equation reads \cite{Becher:2006mr}
\begin{equation}
C_V(-Q^2,\mu) =  \exp\left[ 2S(\mu_h,\mu) - 2a_{\gamma^q}(\mu_h,\mu) \right]
    \left( \frac{-Q^2}{\mu_h^2} \right)^{-a_\Gamma(\mu_h,\mu)}\, C_V(-Q^2,\mu_h) \,,
\end{equation}
where the integrals over the anomalous dimensions are defined as
\begin{equation}\label{RGEsols}
   S(\mu_0,\mu) 
   = - \int_{\mu_0}^\mu\!\frac{d\bar\mu}{\bar\mu}\,\ln\frac{\bar\mu}{\mu_0}\,
    \Gamma_{\rm cusp}^F\big(\alpha_s(\bar\mu)\big) \,, \qquad   
   a_\Gamma(\mu_0,\mu) 
   = - \int_{\mu_0}^\mu\!\frac{d\bar\mu}{\bar\mu}\,
    \Gamma_{\rm cusp}^F\big(\alpha_s(\bar\mu)\big) \,,    
\end{equation}
and similarly for the function $a_{\gamma^q}$. Explicit results for these functions, together with the necessary anomalous dimensions can be found, for example, in Appendix B of \cite{Becher:2010tm}. 


\begin{figure}[t!]
\begin{center}
\begin{tabular}{ll}
\includegraphics[height=0.3\textwidth]{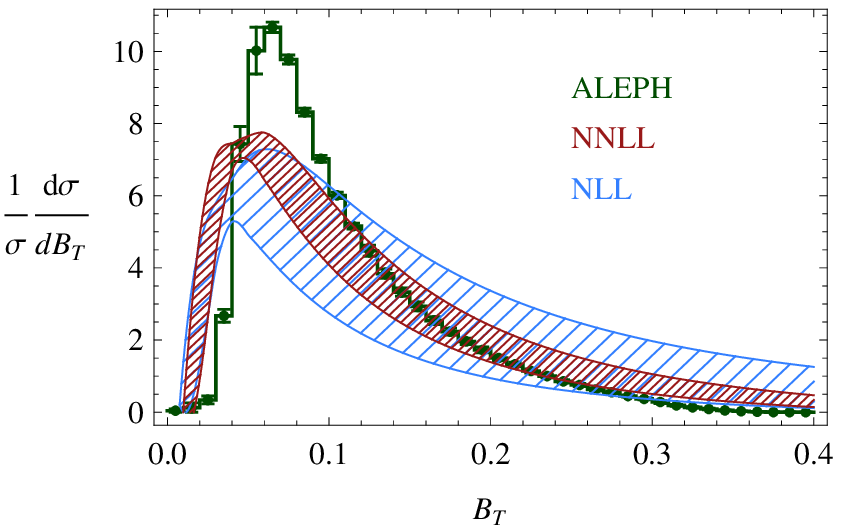} & 
\includegraphics[height=0.3\textwidth]{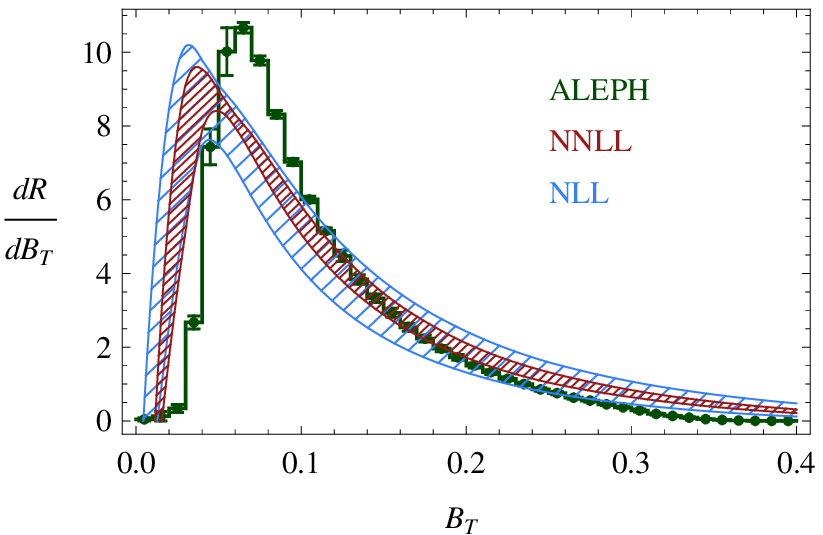} \\
\includegraphics[height=0.3\textwidth]{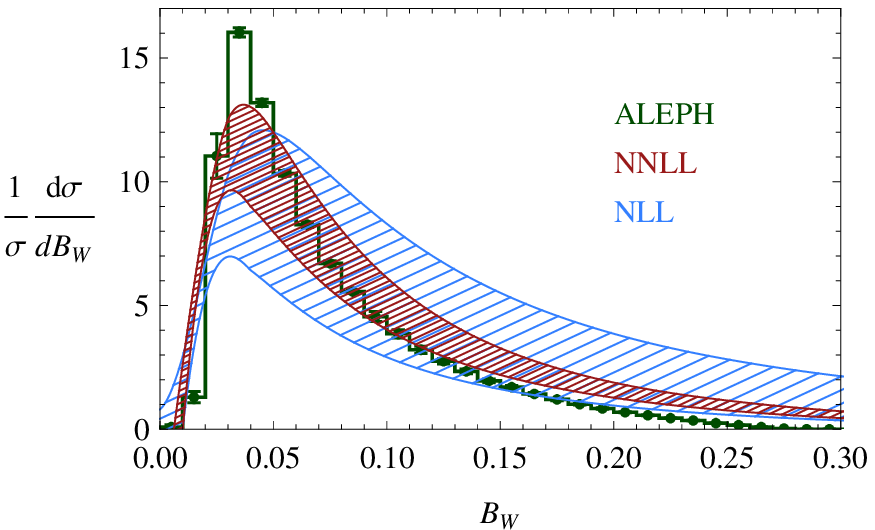} & 
\includegraphics[height=0.3\textwidth]{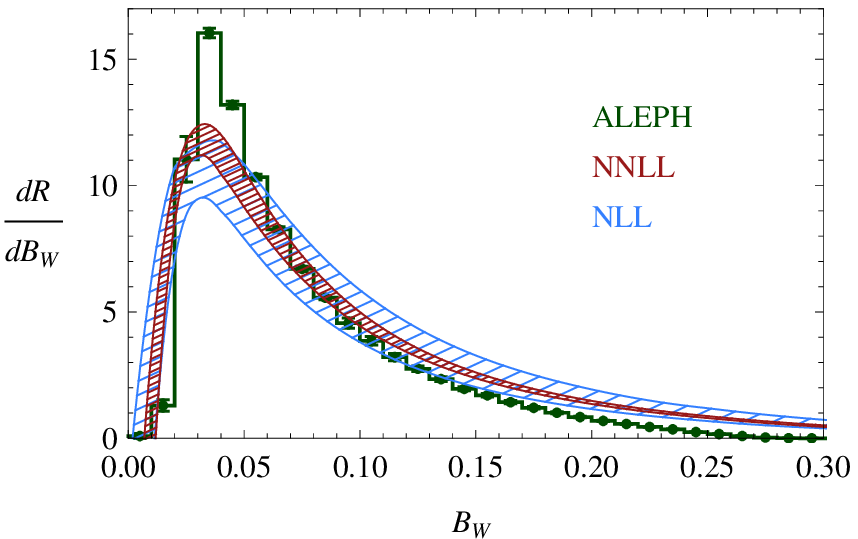} 
\end{tabular}
 \vspace{-0.3cm}
\end{center}
\caption{\label{NNLLnumerical}
NLL and NNLL predictions for the total and wide broadening at $Q=M_Z$. As a reference, we also plot the experimental results of the ALEPH collaboration.}
\end{figure}

In Figure \ref{NNLLnumerical}, we show the result obtained from a numerical evaluation of the  differential broadening cross sections at NLL and NNLL. We normalize to the total hadronic cross section, whose perturbative expansion to NLO reads
\begin{equation}\label{eq:sigtot}
\sigma=\sigma_0\left(1+\frac{3C_F\alpha_s}{4\pi} 
\right)\,.
\end{equation}
The NLL and NNLL distributions are normalized to the fixed-order total cross sections at LO and NLO, respectively. We use dimensionless variables $B_{T}=b_T/Q$ and $B_{W}=b_W/Q$, set $Q=M_Z$ and use the PDG value $\alpha_s(M_Z)=0.1184$ for the strong coupling constant \cite{Beringer:1900zz}. 
The upper two plots show the total broadening distribution, the lower ones the wide broadening. The bands in the plots are obtained by varying the scales by a factor two around their default values $\mu=b_T$ (or $\mu=b_W$ for wide broadening) and $\mu_h = Q$. The resulting variations are added in quadrature. The dominant uncertainty comes from varying the low scale $\mu$. At NNLL (NLL), the hard scale variation is around 1\% (15\%). To use our results for a determination of $\alpha_s$, one will have to also include the matching to fixed-order results to obtain reliable predictions at larger values of $B_T$ or $B_W$. Furthermore, one will need to get a handle on hadronization effects, which modify the distributions in the peak region and below. Without accounting for these two effects, a comparison to experimental data is not completely meaningful, nevertheless we include the ALEPH data \cite{Heister:2003aj} in the plots as a reference. Experimental results are also available from DELPHI \cite{Abdallah:2004xe}, L3 \cite{Achard:2004sv} and OPAL \cite{Abbiendi:2004qz}.

The plots on the left and right side in  Figure \ref{NNLLnumerical} show two different ways of setting the scale $\mu$. We can either directly compute the differential cross section using (\ref{finalres}), or we can obtain it as a derivative of the integrated cross section 
\begin{equation} \label{Rsigtot}
R(B_T) = \int_0^{B_T} dB_T\, \frac{1}{\sigma} \frac{d\sigma}{dB_T}.
\end{equation}
For a fixed value of the scale $\mu$ it is trivial to integrate (\ref{finalres}) to obtain a resummed result for $R(B_T)$. The only reason that the numerical results on the left and right side in Figure \ref{NNLLnumerical} differ is the choice of the scale: choosing $\mu\propto b_T=Q\, B_T $ in $R(B_T)$ does not commute with taking the derivative with respect to $B_T$. The differences between the two prescriptions are not negligible. In particular, we observe that the scale uncertainties come out much smaller when working with $R(B_T)$. However, in both cases, we observe that the NNLL bands nicely overlap with the NLL result.

In the traditional resummation literature \cite{Catani:1992jc,Dokshitzer:1998kz}, the differential distribution is obtained as a derivative of $R(B_T)$. An argument in favor of using this prescription is that the distribution is then properly normalized once the matching is included, since the integral over it produces $R(B_T)$ and the resummation effects become small at large values of $B_T$. In fact, in addition to performing the resummation for $R(B_T)$, the resummed logarithms are usually modified such that they vanish at the end-point $B_T^{\rm max}$, for example by replacing \cite{Dokshitzer:1998kz}
\begin{equation}
 \frac{1}{B_T} \to  \frac{1}{B_T}- \frac{1}{B_T^{\rm max}}+1\,.
\end{equation}
Once this is done $R(B_T^{\rm max})=1$. On the other hand, from an effective field theory standpoint, it is more natural to compute the differential spectrum directly (if this is the quantity of interest) and choose the renormalization scale according to the physical scales in the problem. Determining the spectrum by differentiating $R(B_T)$, one effectively takes the difference of two large quantities, which are evaluated with slightly different scale choices. It has been argued that this leaves spurious contributions in the difference and should be avoided \cite{Abbate:2010xh}.

\section{Fixed-order expansion\label{sec:fixed}}

\begin{figure}[t!]
\begin{center}
\begin{tabular}{cc}
\begin{tabular}{r} 
 \includegraphics[width=0.47\textwidth]{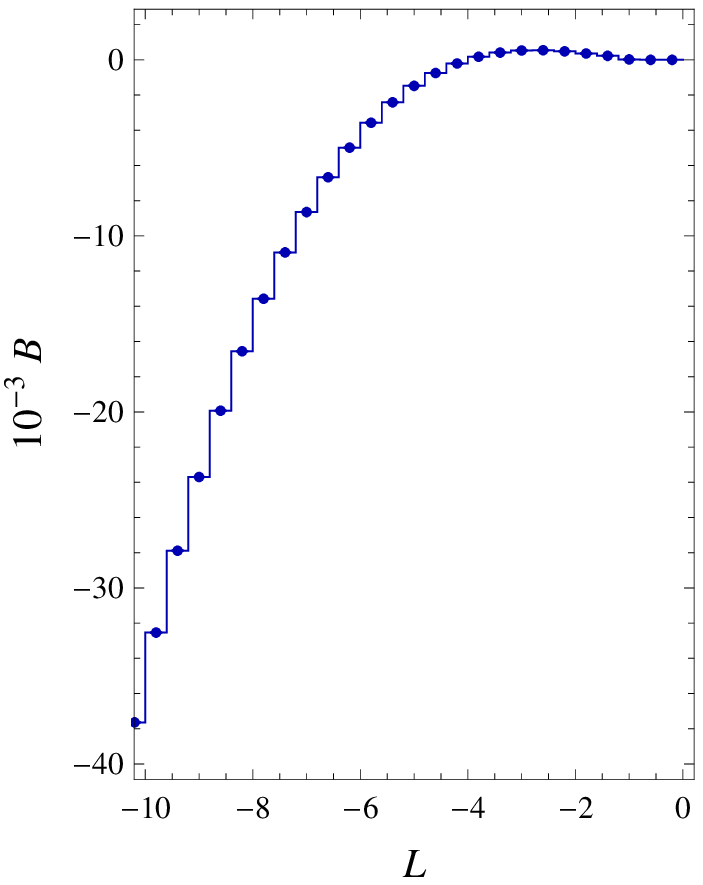}\\
 \includegraphics[width=0.45\textwidth]{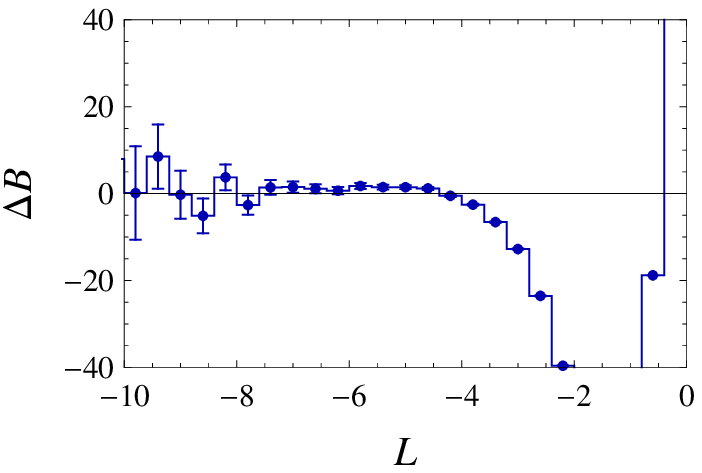} 
 \end{tabular}
 \begin{tabular}{c}
 \includegraphics[width=0.45\textwidth]{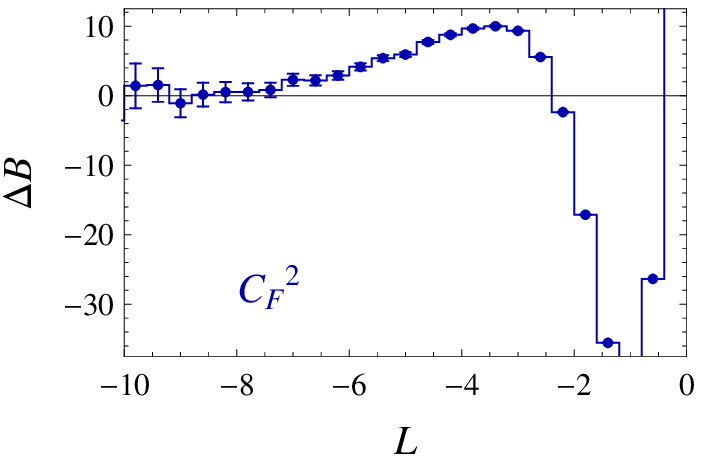} \\
 \includegraphics[width=0.45\textwidth]{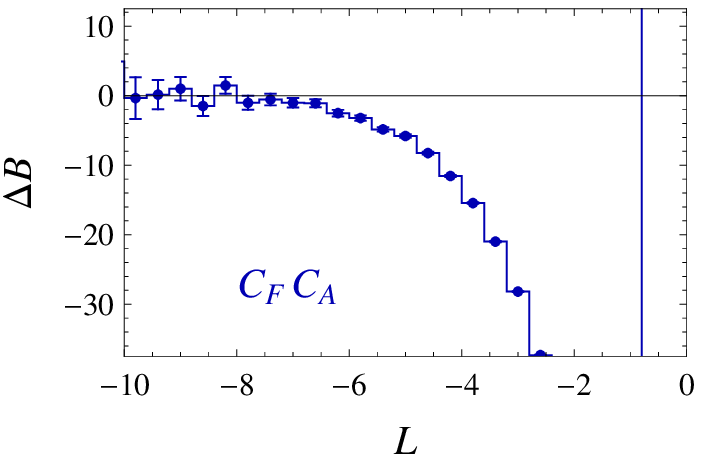} \\
 \includegraphics[width=0.45\textwidth]{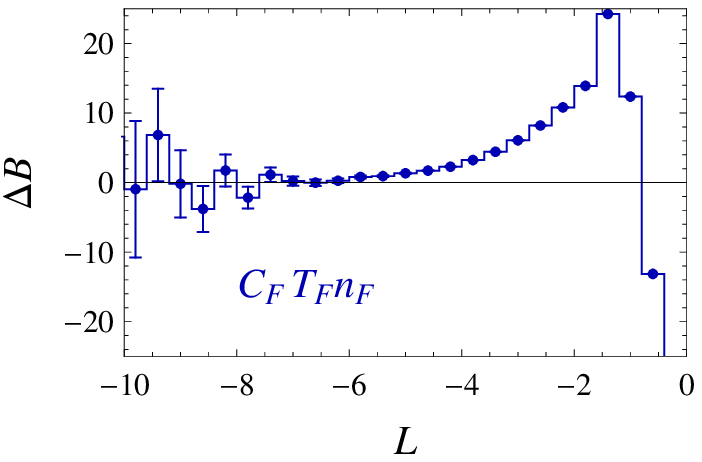} 
 \end{tabular}
 \end{tabular}
 \vspace{-0.5cm}
\end{center}
\caption{\label{eventtwo}
Numerical comparison to results obtained using the {\sc Event2} generator. The upper left panel shows the coefficient $B$, as computed by {\sc Event2}, and the lower panel the difference $\Delta B$ to our result. The difference must go to zero for large negative values of the logarithm $L=\ln B_T$. The right panel shows the contribution of individual color structures in $\Delta B$.}
\end{figure}

It is instructive to perform the fixed-order expansion of the resummed result. To this end, we write the expansion in the form
\begin{equation}\label{exp2}
   \frac{B_T}{\sigma_0}\,\frac{d\sigma}{dB_T} 
   = \frac{\alpha_s(Q)}{2\pi}\,A(B_T) + \bigg( \frac{\alpha_s(Q)}{2\pi} \bigg)^2 B(B_T) + \bigg( \frac{\alpha_s(Q)}{2\pi} \bigg)^3 C(B_T) \,.
\end{equation}
The coefficients $A$ and $B$ can be computed numerically using the {\sc Event2} generator \cite{Catani:1996vz}. The NNLO coefficient $C$ has been obtained in \cite{GehrmannDeRidder:2007hr,Weinzierl:2009ms}. Note that the expansion coefficients are normalized to the tree-level rate. In addition to the differential spectrum, we also consider the integrated rate $R(B)$ for $B=B_T$ or $B=B_W$, which was defined in (\ref{Rsigtot}).  We write the expansion of the logarithmic part of $R(B)$ in the form
\begin{equation}\label{Rexp}
R_L(B)= \left(1+\sum_{k=1}^{\infty} \, C_k \left(\frac{\alpha_s(Q)}{2\pi}\right)^k \right) 
\exp\left[ \sum_{i=1}^{\infty}\sum_{j=1}^{i+1} G_{ij} \, \left(\frac{\alpha_s(Q)}{2\pi}\right)^i \,\ln^j\!\left(\frac{1}{B}\right) \right]\,.
\end{equation}
The quantity $R_L(B)$ is the leading term in the expansion of $R(B)$ around small $B$. In the following, we will give explicit results for the expansion coefficients and will compare our result for $R_L(B)$ with numerical evaluations of $R(B)$ at small values of the broadening. A NLL computation determines the one-loop coefficients $G_{12}$ and $G_{11}$ as well as the two-loop coefficients $G_{23}$ and $G_{22}$. Our computation gives, for the first time, also the coefficient $G_{21}$. For the total broadening, we obtain
\begin{align}\label{G21coeff}
G_{21}&= C_F^2 \left(40 \zeta_3-\frac{45}{2}+4 \pi
   ^2+\frac{128 }{3} \ln ^32+48 \ln ^22-\frac{32}{3} \pi ^2 \ln 2\right)\nonumber\\
   &\hspace{0.5cm}+ C_F C_A \left(-\frac{172 }{3}G+8 \zeta_3+\frac{109}{18}+\frac{11 \pi ^2}{3}+\frac{64 }{3}\ln
   ^32+\frac{88}{3} \ln ^22+\frac{256}{9} \ln 2-4 \pi ^2 \ln
   2\right) \nonumber\\
   &\hspace{0.5cm} +C_F n_f T_F   \left(\frac{56 }{3}G-\frac{10}{9}-\frac{4 \pi ^2}{3}-\frac{32}{3}  \ln
   ^22-\frac{80 \ln 2}{9}\right) \, .
\end{align}
In addition to $\zeta$-values, the result also contains Catalan's constant $G\approx 0.915966$.  To obtain the coefficient, one needs to evaluate moments of the remainder function $w(\tau,z,\mu) $ and the two-loop anomaly exponent $F_B(\tau,z,\mu) $. These were given in equations (\ref{wmoments}) and (\ref{dBmoment}). With the constant $G_{21}$ in place, the two-loop coefficient $B(B_T)$ is known analytically in the limit of small $B_T$ and can be compared to the numerical results of {\sc Event2}. This is done in Figure \ref{eventtwo}, where we show the coefficient as a function of $L=\ln B_T$. At small values of $B_T$, the difference between the logarithmic terms and the full result must go to zero. We find that this is indeed the case within the numerical uncertainties, see Figure  \ref{eventtwo}. Instead of comparing the analytical result to {\sc Event2}, we can also use the program to extract the coefficient $G_{21}$ numerically. The values obtained in this way have an accuracy of about two per cent and agree with our result (\ref{G21coeff}) within their numerical uncertainty, which supports the correctness of our computation and the validity of the resummation formula. The same is true for wide broadening; the relevant expansion coefficients are given in Appendix \ref{sec:expcoeff}. Our values for $G_{21}$ also agree well with the numerical results obtained earlier in \cite{Dokshitzer:1998kz}. 

\begin{figure}[t!]
\begin{center}
 \includegraphics[width=0.95\textwidth]{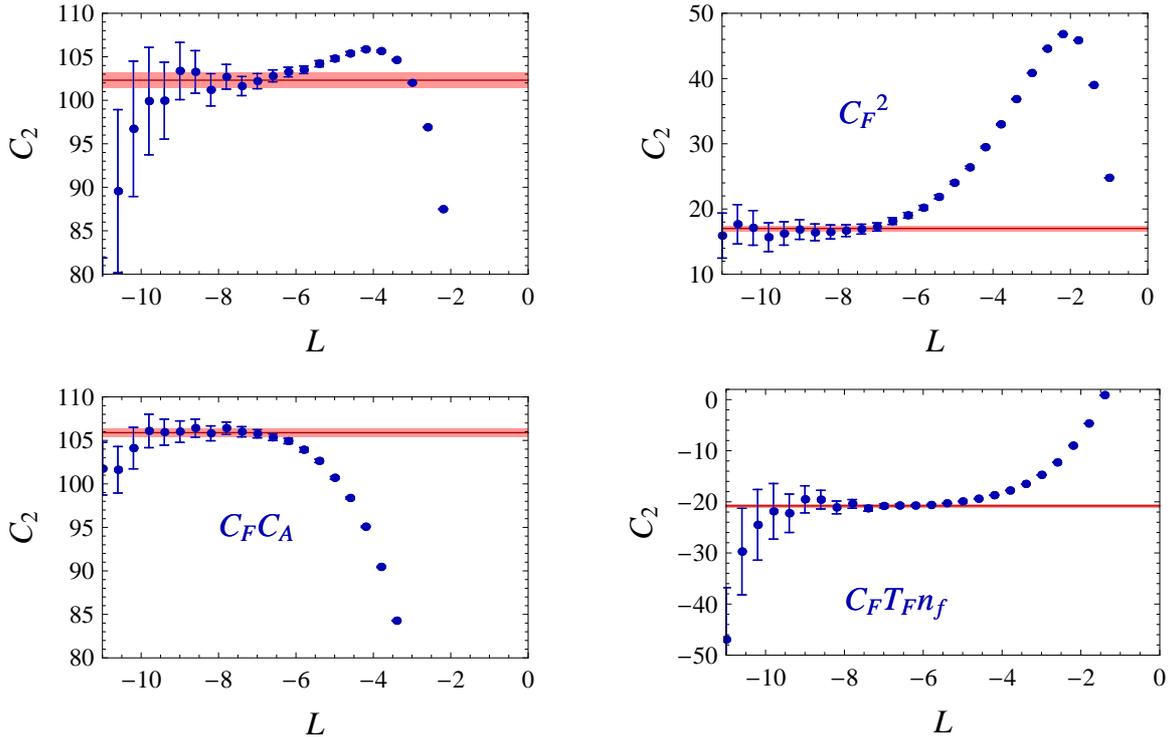}
 \vspace{-0.5cm}
\end{center}
\caption{\label{Ctwo}
Determination of the two-loop constant $C_2$ for total broadening. The first plot is the full coefficient, the remaining ones show the individual color structures.}
\end{figure}

\begin{figure}[t!]
\begin{center}
 \includegraphics[width=0.9\textwidth]{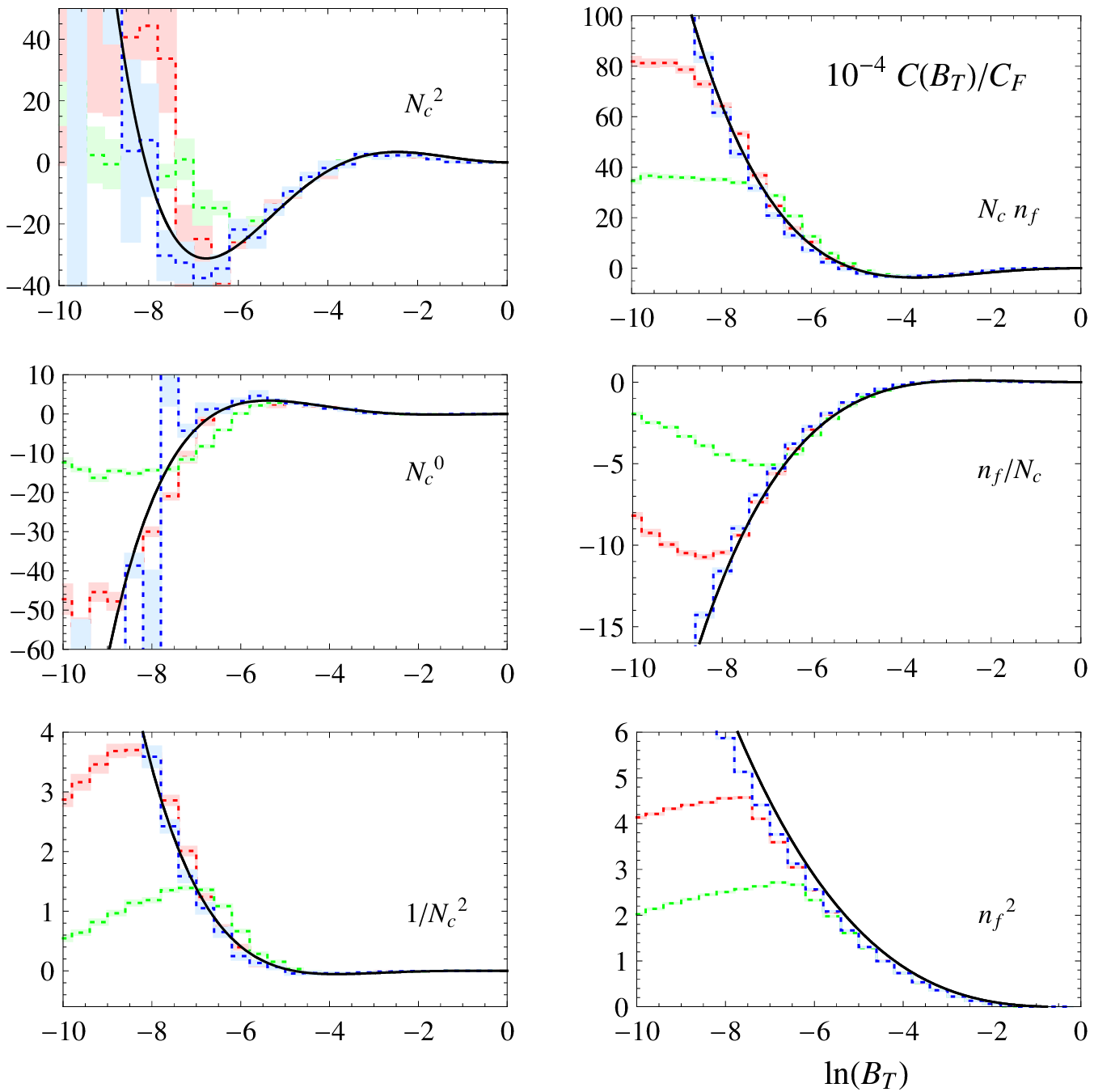}
 \vspace{-0.5cm}
\end{center}
\caption{\label{eeradthree}
Comparison with {\sc eerad3} \cite{GehrmannDeRidder:2007hr} for total broadening for the individual color structures. The {\sc eerad3} code uses an internal infrared cutoff $y_0$ for numerical stability. The three histograms correspond to $y_0=10^{-5}$ (green), $10^{-6}$ (red) and $10^{-7}$ (blue). Our result is the black line. Larger values of the cutoff $y_0$ lead to deviations at small $\ln(B_T)$.}
\end{figure}

Using our analytical results, we can extract the constant $C_2$, which is the last unknown two-loop ingredient. To do so, we compute $R(B)$ in the form
\begin{equation}\label{RB}
R(B) = 1  - \int_B^{B_{\rm max}} dB\, \frac{1}{\sigma}\,\frac{d\sigma}{dB}\,.
\end{equation} 
The integral on the right-hand side of (\ref{RB}) can be evaluated numerically using {\sc Event2}, taking into account the normalization to the total cross section $\sigma$ to NLO, which was given in (\ref{eq:sigtot}). We then consider the difference between the full result $R(B)$ and its logarithmic part $R_L(B)$. Since everything except the constant $C_2$ is known, the constant immediately follows from the requirement that the difference must vanish in the limit $B\to 0$. Figure \ref{Ctwo} shows the difference as a function of the logarithm as well as the value of $C_2$ we extract from it. For the total broadening, we obtain
 \begin{align}\label{eq:C2total}
 C_2 = (105.9 \pm 0.5)_{C_F C_A} +  (17.0 \pm 0.4)_{C_F^2} +  (-20.8 \pm 0.3)_{C_F T_F n_f} = 102.3 \pm 0.9 \,,
 \end{align}
 and the result for the wide broadening is
\begin{align}\label{eq:C2wide}
 C_2 = (128.5 \pm 0.6)_{C_F C_A} +  (-0.41\pm 0.27)_{C_F^2} +  (-15.0 \pm 0.2)_{C_F T_F n_f} = 113.2 \pm 0.4 \,.
 \end{align} 
 The number for the total contribution has been obtained by fitting the sum of all color structures. Within uncertainties, it agrees with the result obtained from adding the fit results of the individual color structures. Figure \ref{Ctwo} makes it clear that it is somewhat delicate to extract these numbers. On one hand, one wants to make the broadening as small as possible to avoid power suppressed terms which contribute to $R(B)-R_L(B)$. On the other hand, one cannot make it too small because the numerics become unstable at very low $B$. To be able to reach very low values of $B$, we have run {\sc Event2} at quadruple precision and have generated $10^8$ events with a very low cutoff of $10^{-16}$. To obtain the above values and uncertainties, we have performed $\chi^2$ fits using all   fit intervals in the range $L\in (-11.8,-4.6)$ containing at least six fit points. We then select the 12 fits with the lowest $\chi^2$-value and use the spread among these as an estimate of the systematic uncertainty due to the choice of the fit range. The statistical and systematic uncertainties are of similar size and have been added in quadrature for the error estimates in (\ref{eq:C2total}) and (\ref{eq:C2wide}).

\begin{figure}[t!]
\begin{center}
 \includegraphics[width=0.9\textwidth]{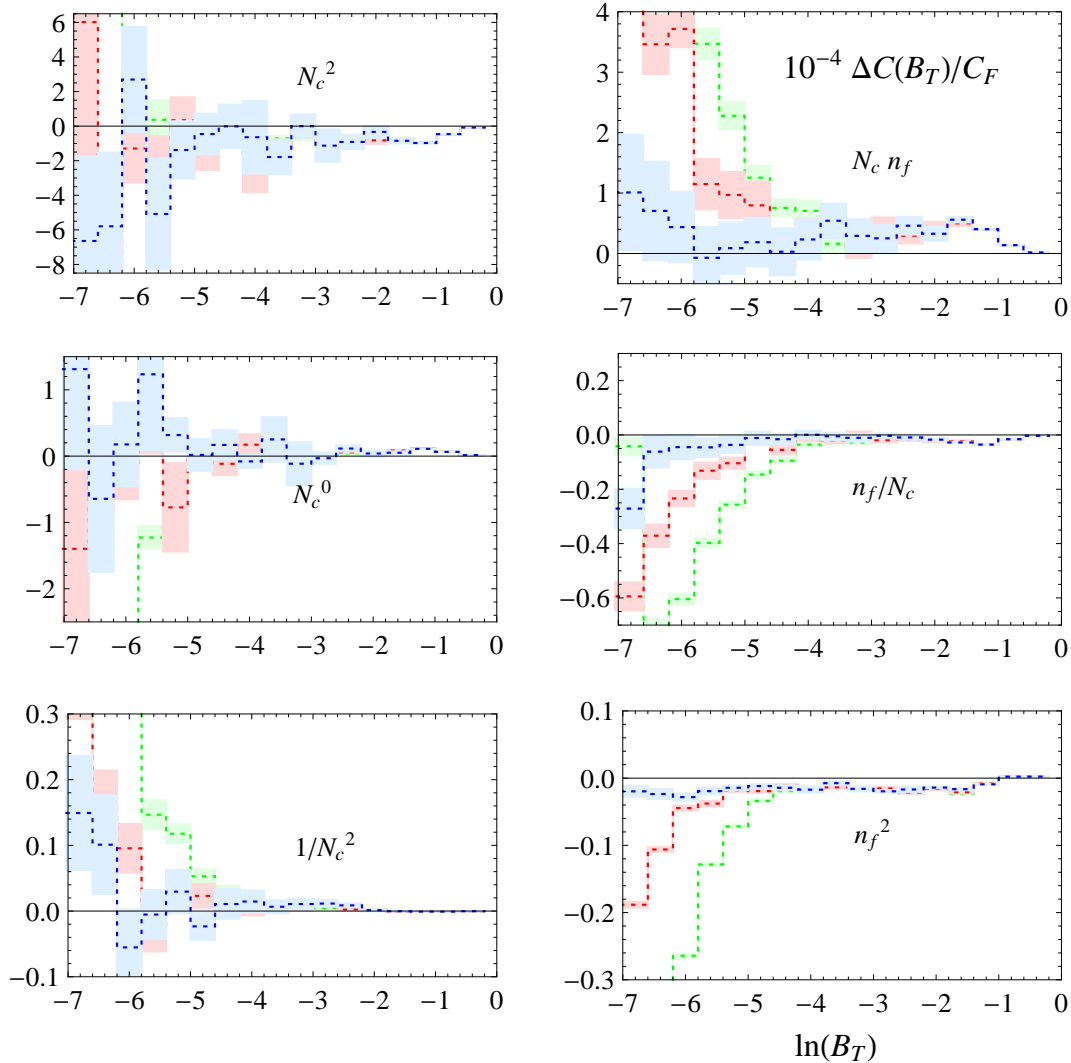}
 \vspace{-0.5cm}
\end{center}
\caption{\label{eeradthreeDiff}
Same as Figure \ref{eeradthree}, except that we plot the difference between {\sc eerad3}  and our prediction for the singular terms. The difference must become constant at small broadening. Up to cutoff effects, this is indeed the case.}
\end{figure}

It is also interesting to compare our results to the numerical NNLO predictions obtained from {\sc eerad3} \cite{GehrmannDeRidder:2007hr}.\footnote{We thank Thomas Gehrmann for providing us with the results.} These provide the coefficient $C(B_T)$ in (\ref{exp2}), which is plotted in Figure \ref{eeradthree}. The comparison serves two purposes: first of all, it allows us to check that the resummation formula, by verifying that we correctly obtain the numerically dominant ${\mathcal O}(\alpha_s^3)$  terms at small values of the broadening. Secondly, it tests the numerical stability of  {\sc eerad3} at small values of the broadening and the sensitivity to the internal cutoff. To avoid numerical instabilities, the generator imposes a cut on the phase-space variables \cite{GehrmannDeRidder:2007jk}. The observables are independent of the value of the cut parameter $y_0$, if it is chosen small enough. By comparing to the numerical results at very small broadening, we can check the residual sensitivity to the choice of $y_0$. We show three different histograms in Figure \ref{eeradthree}, corresponding to three different choices of $y_0$. For the larger values of $y_0$ one observes dramatic deviations at small $B_T$, while we observe nice agreement for the smallest choice $y_0=10^{-7}$ of the cutoff parameter.

Our NNLL computation does not fully determine the logarithmic part at NNLO. It allows us to get the coefficients $G_{34}$, $G_{33}$ and $G_{32}$, which are given in Appendix \ref{sec:expcoeff}, but the coefficient $G_{31}$ is beyond the accuracy of our computation. As a consequence, the difference $\Delta C(B_T)$ between our result and the numerical evaluation using {\sc eerad3} does not vanish, but becomes constant at small broadening. We plot the difference $\Delta C(B_T)$ in Figure \ref{eeradthreeDiff}. Within numerical uncertainties the expected behavior is indeed observed, except at very low $B_T$, where the cutoff effects are non-negligible, as can be seen by comparing the results obtained with different cutoffs. We have performed the same numerical comparison for wide broadening and find consistent results also in this case. 

\section{Conclusion\label{sec:conclusion}}

In this paper, we have extended the resummation for the jet broadening $e^+e^-$ event-shape variables to NNLL accuracy. Our analysis is based on the all-order factorization theorem we obtained earlier \cite{Becher:2011pf}. The factorization for broadening suffers from a collinear anomaly, and to obtain NNLL accuracy, we have computed the anomaly exponent to two-loop order and the jet and soft functions to one-loop accuracy.

The presence of the anomaly makes the computation subtle, since one needs to introduce an additional regulator beyond dimensional regularization to make the individual jet and soft functions well-defined. We recently conceived a convenient way of introducing the additional regularization, which is based on a modification of phase space, and respects gauge invariance and the factorization structure of the effective theory \cite{Becher:2011dz}. An alternative scheme was proposed in \cite{Chiu:2011qc,Chiu:2012ir}. Together with the recent work \cite{Gehrmann:2012ze},  our paper provides the first example of a full two-loop computation in SCET for an observable which is sensitive to soft recoil and the results demonstrate that our regularization method works in practice. 

The anomaly coefficient was extracted from the two-loop broadening soft function computed at fixed recoil momentum against the collinear radiation. The associated constraint on the two-emission phase space is quite complicated. Despite the complexity of the relevant phase-space integrals, we have managed to perform the computation analytically. In addition to polylogarithmic functions, we find that the result also involves elliptic integrals. As a consequence, the expansion coefficients for jet broadening do not only involve $\zeta$-values, but also $\beta$-values (where $\beta$ is the Dirichlet $\beta$-function) such as Catalan's constant. Other examples of perturbative computations involving elliptic integrals can be found for instance in \cite{Bailey:2008ib,Czakon:2008ii}. 

In addition to numerically verifying our analytical results for the diagrams, we have also compared the expansion of our resummed result in $\alpha_s$ to the predictions from fixed-order event generators. Running  {\sc Event2} with high statistics and a low cutoff, we have checked that we reproduce the numerical result for the wide and total broadening distributions at  ${\mathcal O}(\alpha_s^2)$ at small values of these parameters. Comparing to {\sc eerad3}, we have performed the same check at ${\mathcal O}(\alpha_s^3)$. Also in this case, we observe the expected behavior.

Determinations of the strong coupling constant from LEP event-shape data typically use about half a dozen different event shapes, among them thrust, heavy jet mass, total and wide broadening \cite{Heister:2003aj,Abdallah:2004xe,Achard:2004sv,Abbiendi:2004qz,Dissertori:2007xa, Dissertori:2009ik}. Statistically, there is little gain in using several observables since all of them are based on the same collider events. However, the consistency among the values extracted from the different observables provides an important cross check on systematic effects and on the reliability of the theoretical uncertainty estimate. With our results for broadening, higher-log resummation is now available for all four of the event shapes mentioned above. It will be interesting to perform an extraction of $\alpha_s$ based on these results and to check whether the low value of $\alpha_s$ obtained from a global fit to thrust data based on N$^3$LL resummation is confirmed \cite{Abbate:2010xh}.

\vspace{4mm}\noindent
{\em Acknowledgements:\/}
We are grateful to Xavier Garcia i Tormo, Thomas Gehrmann, Tobias Huber, Christian Lorentzen, Matthias Neubert and Gavin Salam for useful discussions. The work of T.B.\ is supported by the Swiss National Science Foundation (SNF) under grant 200020-140978 and the Innovations- und Kooperationsprojekt C-13 of the Schweizerische Universit\"atskonferenz (SUK/CRUS).

\begin{appendix}

\section{Expansion coefficients\label{sec:expcoeff}}

Here we give the NNLL expansion coefficients defined in (\ref{Rexp}) up to three-loop order. For total broadening, the coefficients read
\begin{align}
 C_1 & = \left(\pi^2-\frac{17}{2} \right) C_F \,, \nonumber\\
 G_{12}&= -4 C_F \,,  \nonumber\\
 G_{11}&= 6 C_F \,,\\
 G_{23}&= \frac{32}{9} C_F n_f T_F-\frac{88}{9} C_A C_F \,,  \nonumber\\
 G_{22}&= \left(\frac{2 \pi ^2}{3}-\frac{35}{9}\right) C_A C_F
+\frac{4}{9} C_F n_f T_F+\left(-\frac{8 \pi ^2}{3}-32 \ln^22\right) C_F^2\,.  \nonumber
\end{align}
The coefficient $G_{21}$ was given in (\ref{G21coeff}). The three-loop coefficients are
\begin{align}
 G_{34}&= \frac{176}{9} C_A C_F n_f T_F-\frac{242}{9} C_A^2 C_F-\frac{32}{9} C_F n_f^2 T_F^2 \,,  \nonumber \\
 G_{33}&=  \left(-\frac{32 \pi ^2}{27} +\frac{3040}{81} \right) C_A C_F n_f T_F+\left(\frac{88 \pi ^2}{27}-\frac{4942}{81}\right) C_A^2 C_F \nonumber\\
&\hspace{0.5cm}  +\left(-\frac{176 \pi ^2}{9}-\frac{704 \ln^22}{3}\right) C_A C_F^2-\frac{352}{81} C_F n_f^2 T_F^2 \nonumber \\
&\hspace{0.5cm}+\left(\frac{64 \pi^2}{9}+\frac{256 \ln^22}{3} +\frac{16}{3}\right) C_F^2 n_f T_F+ \left(\frac{176 \zeta _ 3}{3}-128 \ln^32+\frac{64}{3} \pi ^2 \ln2\
\right) C_F^3 \,, \\
 G_{32}&=-41.46 \,C_A C_F n_f T_F-200.62\, C_A C_F^2+78.75\, C_A^2 C_F+97.09 \,C_F^2 n_f T_F \nonumber \\ 
 &\hspace{0.5cm} +9.103\, C_F n_f^2 T_F^2-166.92\, C_F^3\,.  \nonumber
\end{align}
We only list the numerical value for $G_{32}$, but by computing higher logarithmic moments of the remainder and the anomaly functions, it could be obtained analytically from our results.

For wide broadening, the leading one-loop coefficients as well as the leading-logarithmic higher-order coefficients agree with the total broadening. For the remaining coefficients, we list the difference to the total broadening coefficients, for which we obtain
\begin{align}
  \Delta G_{22}&= \frac{8\pi ^2}{3}  C_F^2 \,,\nonumber \\
 \Delta G_{21}&=  \left(32 \zeta _ 3-4 \pi ^2\right) C_F^2 \,,\nonumber \\
\Delta G_{33}&=  \frac{176\pi^2}{9}  C_A C_F^2-\frac{64\pi ^2}{9}  C_F^2 n_f T_F-128 \zeta _ 3 C_F^3\,,  \\
 \Delta G_{32}&= \left(-\frac{8 \pi ^4}{9}+352 \zeta _ 3-\frac{256 \pi ^2}{27}\right) C_A C_F^2+\left(-128 \zeta _ 3+\frac{128 \pi ^2}{27}\right) C_F^2 n_f T_F  \nonumber \\ 
 &\hspace{0.5cm} + \left(-\frac{208 \pi ^4}{45}+288 \zeta _ 3+\frac{128}{3} \pi ^2 \ln^22\right) C_F^3 \,.\nonumber
\end{align}

\section{Numerical evaluation of the jet and soft functions \label{sec:numerical}}

As a check of our results, we have independently evaluated the diagrams numerically. The numerical evaluation is done directly in Laplace-Fourier space. The relevant phase-space integrals suffer from soft and collinear divergences, which we extract using a combination of subtraction and sector decomposition methods. 

\subsection{Jet function\label{sec:numerical:jet}}

We start with the self-energy diagram and compute its Laplace and Fourier transform 
which yields
\begin{equation}
 \widetilde{\cal J}^{(1a)}_L(\tau,x_\perp,\mu) =\alpha_s C_F \frac{2^{1+2\epsilon}}{\pi^{2-2\epsilon}}(1-\epsilon)\tilde{\mu}^{2\epsilon} \frac{1}{(2\pi)^{d-2}}\,  {I}(\tau,x_\perp,\mu)\, ,
\end{equation} 
where the relevant phase-space integral is given by
\begin{equation}
{I}(\tau,x^\perp,\mu) = \int \!d^dq\, \delta(q^2) \theta(q^0) \int\! d^dk\, \delta(k^2)  \theta(k^0)   \delta(Q-q_- - k_-)   \frac{k_-}{2k\cdot q}\, e^{-i (k_\perp+q_\perp) \cdot x_\perp}  \,e^{-\frac{\tau}{2}(k + q) } \,,\label{Ifun}
\end{equation}
and we use the notation $k=|k_\perp| = \sqrt{-k_\perp^2}$ and $q=|q^\perp|$.
This is the same as expression (\ref{self}), except that the delta functions have turned into exponentials because of the Laplace and Fourier transforms.

The jet function only depends on the magnitude of $x_\perp$, but not its direction. We can therefore average over the direction of $x_\perp$ relative to the total momentum $k_\perp+q_\perp$ and replace 
\begin{equation}
 e^{-i (k_\perp+q_\perp) \cdot x_\perp} \to \frac{\Omega_{d-3}}{\Omega_{d-2}}  \int_0^\pi d\phi  \,\sin^{d-4}\phi \,e^{-i |k_\perp+q_\perp|  x \cos\phi}
\end{equation}
inside the integral, with $x=|x^\perp|$. We then use light-cone integration variables and integrate over the $k_+$, $q_+$ and $q_-$ components, which leads to
\begin{align}
{I}(\tau,x^\perp,\mu) &=\frac{ \Omega_{d-3}^2}{4} \int_0^Q dk_-\int_0^\infty dk \,k^{d-3}  \int_0^\infty dq\, q^{d-3} \int_0^\pi d\theta \sin^{d-4}\theta  \int_0^\pi d\phi \sin^{d-4}\phi \nonumber \\
&\hspace{2cm}\frac{1}{q_-}\frac{1}{k_+ q_- + q_+ k_- - 2k q \cos\theta} e^{-\frac{\tau}{2}(k + q)+ i  x\sqrt{k^2+q^2+2 \cos\theta k q } \cos\phi}\,,
\end{align}
with $k_+ = k^2/k_-$ and $q_+ = q^2/q_-$ and $q_- = Q-k_-$ due to the three delta functions in (\ref{Ifun}).

We now eliminate $x$ in favor of the dimensionless variable $z=\frac{2x}{\tau}$ and introduce new integration variables $u$, $v$ and $\eta$ ($\bar{u} = 1-u$ and $\bar{v} = 1-v$): 
\begin{align}\label{variables}
k_- &= Q u\, ,& k &= \frac{\eta \bar{v}}{\tau \bar{u} }\, ,& q &= \frac{\eta v}{\tau u}  \,.
\end{align}
After the variable change, the integration over $\eta$ can be performed, which yields the result
\begin{align}
I(\tau,x^\perp,\mu)  &= 4^{d-4} \Omega_ {d-3}^2 \Gamma (2 d-6) \tau ^{6-2 d}\,\int_0^1 \frac{du}{\bar{u}} \int_0^1 dv   \int_0^\pi d\theta \sin ^{d-4}\theta  \,
\frac{\left( u \bar{u} v \bar{v}\right)^{d-3} }{v^2+\bar{v}^2-2 v \bar{v} \cos\theta}\nonumber \\
   &\hspace{1cm}
\int_0^\pi d\phi \sin ^{d-4}\phi   \left( u \bar{v} +\bar{u} v-i z \cos\phi \sqrt{u^2 \bar{v}^2 +\bar{u}^2 v^2+2 u v \bar{u} \bar{v} \cos\theta}\right)^{6-2 d}\,.
\end{align}

Without the term on the second line, we can perform the integral over $v$ and $\theta$ analytically in $d=4-2\epsilon$ dimensions
\begin{equation}\label{kernelint}
\int_0^1 dv   \int_0^\pi d\theta \sin ^{d-4}\theta  \frac{ \left(v \bar{v}\right)^{d-3}}{v^2+\bar{v}^2-2 v \bar{v} \cos\theta} = - \frac{4^{\epsilon -1} \sqrt{\pi }\, \Gamma \left(\frac{3}{2}-\epsilon
   \right)^2}{(1-\epsilon) \epsilon\,  \Gamma \left(\frac{3}{2}-2 \epsilon \right)} = -\frac{\pi}{8\epsilon} +\dots\,.
\end{equation}
The divergence arises from the collinear singularity at the point $v=\frac{1}{2}$ and $\theta=0$. This is the only singularity of the original integral, and we can isolate it by rewriting it in the form
\begin{equation}\label{subtraction}
I(\tau,x^\perp,\mu) = \int_0^1 dv   \int_0^\pi d\theta \sin ^{d-4}\theta  \frac{ v^{d-3} \bar{v}^{d-3}}{v^2+\bar{v}^2-2 v \bar{v} \cos\theta}\left\{ \left[ f(v,\theta) - f(\mbox{$\frac{1}{2}$},0)  \right] +  f(\mbox{$\frac{1}{2}$},0)\right\}.
\end{equation}
The subtracted part in square brackets is finite for $d=4$ and can be evaluated numerically. In the remainder, the $v$ and $\theta$ integrations can be evaluated in $d$ dimensions using (\ref{kernelint}). Once this is done, the integral can be expanded around $d=4$ on the level of the integrand and the remaining integration can be performed numerically.

The jet-function diagrams involving Wilson-line emissions suffer from soft divergences in addition to the collinear singularity present in the self-energy diagram. In the parameterization (\ref{variables}), the soft singularities appear when $u$ and $v$ go to zero and are regularized analytically. To isolate the two types of divergences, we split the $v$ integration as
\begin{equation}
\int_0^1 d v = \int_0^{v_{\rm cut}} dv +  \int_{v_{\rm cut}}^1 dv\,.
\end{equation}
As long as $0< v_{\rm cut} < 1/2$, the first integral contains only soft divergences. The second part only suffers from the collinear divergence and can be evaluated numerically using the same technique as in the case of the self-energy integral. To compute the soft part, we first rescale $v$ so that it runs again from $0\dots 1$ and then use sector decomposition \cite{Hepp:1966eg,Roth:1996pd,Binoth:2000ps}
 and split the integration over $u$ and $v$ into the sectors  $u<v$ and $u> v$. After rescaling $u \to v \,u$ in the first sector and $v \to u \,v$ in the second, the divergences are disentangled and can be extracted using relations such as
\begin{equation}
v^{-1+\alpha} = \frac{1}{\alpha} \delta(v) + \left[ \frac{1}{v} \right]_+ + \alpha \left[ \frac{\ln v}{v} \right]_+  + {\mathcal O}(\alpha^2) \,.
\end{equation}

\begin{figure}[t!]
\begin{center}
\includegraphics[width=0.4\textwidth]{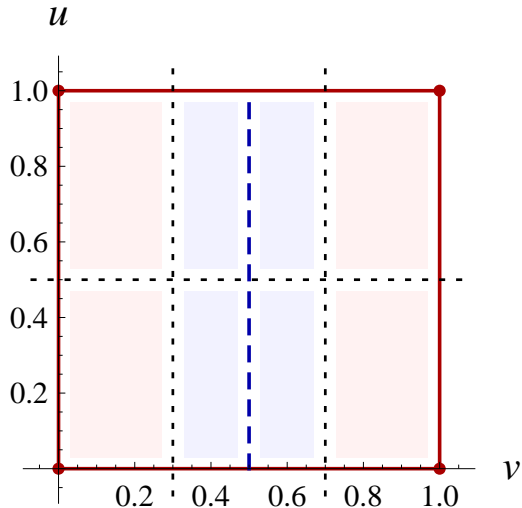} 
\end{center}
\vspace{-0.5cm}
\caption{\label{sectors}
Sector decomposition for the numerical evaluation of the two-loop diagrams. Along the boundary of the integration region (red square) soft singularities occur, while the collinear singularities are located at $v=\frac{1}{2}$ (blue dashed line). We cut the integration into eight patches along the dotted lines.
}
\end{figure}

\subsection{Two-loop soft function  \label{sec:numerical:soft}}

The one-particle cut diagram (h) in Figure \ref{soft:nnlodiagrams} can easily be evaluated analytically directly in Fourier-Laplace space. For our numerical check, we therefore focus on the two-particle  contributions. As in the jet function case, the corresponding integrals suffer from soft and collinear divergences. 

To illustrate the singularity structure of the relevant integrals, let us consider diagram (e). The corresponding amplitude squared has the form
\begin{equation}
|{\cal M}(q,k)|^2 = 2 C_F C_A  g_s^4 \frac{ \left(2
   k_-+q_-\right)}{k_- q_+ \left(k_-+q_-\right)
   \left(k_+ q_-+k_-
   q_+ -2 k\, q \cos\theta\right)}
\end{equation}
and the correction to the broadening is obtained by evaluating the integral 
\begin{multline}
\int \!d^dq\, \delta(q^2) \theta(q^0) \int\! d^dk\, \delta(k^2) \left(\frac{\nu }{k_+}\right)^{\alpha } \left(\frac{\nu}{q_+}\right)^{\alpha } |{\cal M}(q,k)|^2\\
\times  \theta \! \left(k_+-k_-\right) \theta\!\left(q_+-q_-\right) e^{-i (k_\perp+q_\perp) \cdot x_\perp}  \,e^{-\frac{\tau}{2}(k + q) }\,.
\end{multline} 
The exponentials arise from the Laplace-Fourier transform and the $\theta$-functions ensure that both particles are emitted into the left hemisphere. Other than this, the two-particle phase space relevant for the soft sector is completely unrestricted, in contrast to the collinear phase space for the jet function, where the sum of the large momentum components had to add up to the center-of-mass energy $Q$. After integrating over the $k_+$ and $q_+$ components, we parameterize
\begin{align}
k_- &=  \frac{\eta  u \bar{v} w}{\tau}\,, &   
q_- &=  \frac{\eta  \bar{u}  v w}{\tau }\,, & 
|k_\perp|  & =   \frac{\eta  u}{\tau }\,, &
 |q_\perp| &= \frac{\eta  \bar{u}}{\tau } \,.
\end{align}
The integration over $\eta =0 \dots \infty$ can immediately be done.  Also the integration over $w$ can be carried out. Due to the hemisphere constraint, this variable runs from $w=0 \dots \frac1v$ if $v<\frac12$ and from $w=0 \dots \frac1{\bar{v}}$ otherwise. Up to a prefactor, the integral then takes the form
\begin{multline}
\int_0^1 du\, u^{ -1-\alpha -2 \epsilon} \bar{u}^{-1-\alpha -2  \epsilon}   \int_0^1 dv\,  \left(  v^{2 \alpha } \theta\! \left(v-\mbox{$\frac12$} \right)+  \bar{v}^{2 \alpha }\theta\! \left(\mbox{$\frac12$} -v\right) \right)  v^{1-\alpha } \bar{v}^{-\alpha -1} \frac{ \left(v \bar{u}+2 u  \bar{v}\right)    }{v \bar{u}+u \bar{v}} \\
\times \int_0^\pi d\theta \frac{\sin^{-2 \epsilon } \theta  }{v^2 +\bar{v}^2-2 v \bar{v} \cos\theta} \int_0^\pi d\phi \sin ^{-2 \epsilon }\phi \left(1-i z \cos \phi \,  \sqrt{u^2+\bar{u}^2+2 u \bar{u} \cos\theta}\right)^{2 (\alpha +2 \epsilon)}\,.
\end{multline}
The terms in the first line have soft singularities for $u=0$ and $u=1$ as well as for $v=1$ (for other diagrams, also $v=0$ is singular). For $v=\frac12$, the $\theta$-integral in the second line produces the collinear singularity shown in (\ref{kernelint}). As in the computation of the jet function, we introduce a parameter $v_{\rm cut}$ with $0<v_{\rm cut} <\frac12$ to separate the soft and collinear regions. We then split the integration region into eight sectors as shown in Figure \ref{sectors}. In the four red corner areas, we disentangle the overlapping singularities in $u$ and $v$ using sector decomposition. In the blue collinear regions, we can either work with subtractions as in (\ref{subtraction}), or we can disentangle the singularities in the $v$ and $\theta$ integration using sector decomposition. We prefer the latter, since the subtraction technique becomes cumbersome for the self-energy diagrams (f) and (g), which involve the collinear denominator $v^2 +\bar{v}^2-2 v \bar{v} \cos\theta$ to the second power. For the numerical evaluation, it is advantageous to combine the individual diagrams, since the sum of the diagrams is typically less singular than the individual ones. This is in particular the case for the self-energy diagrams. Working with the sum of the diagrams as given in Appendix \ref{sec:softmatrix}, we manage to perform the numerical evaluation with a relative accuracy of $10^{-6}$ or better.

\section{Soft-gluon matrix element\label{sec:softmatrix}}

We write the squared matrix element describing the emission of two soft gluons in the form
\begin{multline}
|{\cal M}_{2g}(k,q)|^{2} = g_s^4\, \Big[ C_F^2\, {\cal A}(k,q) + C_F C_A \, {\cal N}(k,q)  \\
+  C_F \left( T_F n_f -C_A \right) \, {\cal S}_1(k,q) + C_F \left( T_F n_f- (1-\epsilon) C_A/2 \right) \, {\cal S}_2(k,q)    \Big]\,.
\end{multline}
The relevant diagrams are shown in Figure \ref{soft:nnlodiagrams}. The individual pieces of the amplitude squared are
\begin{align}
{\cal A}(k,q) &= \frac{8}{k_- k_+ q_- q_+} \,,\nonumber \\
{\cal N}(k,q) &=\frac{1}{k_- k_+ q_- q_+}\left[ \frac{k_+ q_-+k_- q_+}{k\cdot q}-\frac{2 \left(k_- k_++q_- q_+\right)}{\left(k_-+q_-\right) \left(k_++q_+\right)} \left (1-\frac{k_+ q_-+k_- q_+}{2 k\cdot q} \right)-2 \right]  \,, \nonumber\\
{\cal S}_1(k,q)  & = \frac{4}{ k\cdot q \left(k_-+q_-\right) \left(k_++q_+\right)}\,, \\
{\cal S}_2(k,q)  & = -\frac{2 \left(k_+ q_--k_- q_+\right)^2}{ (k\cdot q)^2 \left(k_-+q_-\right)^2
   \left(k_++q_+\right)^2} \,.  \nonumber
\end{align}
The abelian part ${\cal A}(k,q) $ is just one half of the square of the one-loop amplitude. The non-abelian piece ${\cal N}(k,q)$ gets contributions from the box diagrams (a), (b) and (c), as well as from diagrams (d) and (e) involving a triple-gluon vertex. The latter involve a propagator denominator $k\cdot q$. The last two structures ${\cal S}_1(k,q)$ and ${\cal S}_2(k,q)$ arise from the self-energy diagrams (f) and (g). 

\section{Results for individual diagrams \label{sec:diagrams}}

In the main text, we have discussed the evaluation of diagram (b) in Figure \ref{soft:nnlodiagrams} in detail. In the following, we summarize our results for the remaining diagrams. Using partial fractioning identities, we first relate diagram (a) to the other two box diagrams giving
\begin{align}
   {\cal S}_L^{(2a)}(b_L,b_R,p_L^\perp, p_R^\perp) =
   \frac{C_F}{C_F - C_A/2}\; \bigg(
   {\cal S}_L^{(2c)}(b_L,b_R,p_L^\perp, p_R^\perp) 
   - {\cal S}_L^{(2b)}(b_L,b_R,p_L^\perp, p_R^\perp)  \bigg)\,,
\end{align}
and this relation carries over to the Laplace-Fourier transformed expression.

We next consider diagram (c), which can be written as a convolution of two NLO integrals. In broadening-momentum space, the exact result reads
\begin{align}
	{\cal S}_L^{(2c)}(b_L,b_R,p_L^\perp, p_R^\perp)
&= \frac{\alpha_s^2 C_F}{4\pi^{3-2\eps}}\, \Big(C_F-\frac{C_A}{2}\Big)\,
   \left(\frac{2\mu^2e^{\gamma_E}}{b_L}\right)^{2\eps} \frac{\Omega_{d-3}}{b_L^3}\;\frac{1}{\alpha^2}\,
   \left(\frac{\nu_+}{b_L}\right)^{2\alpha}  \,
    \delta^{d-2}(p_R^\perp)\;
    \delta(b_R)
\nonumber\\
   &\quad
    \times \; \!(1-y^2)^{-1-2\eps-\alpha}\;
    \frac{\Gamma(1-2\eps)}{\Gamma^2(1-\eps)} \;\,
    {}_2F_1\bigg(\frac12 - \eps, -\eps -\alpha; 1 - \eps; y^2 \bigg)\,.
\end{align}
After Laplace and Fourier transformation this turns into
\begin{align}
&\overline{\cal S}_L^{(2c)}(\tau_L,\tau_R,z_L, z_R) 
   =   
\left(\frac{\alpha_s}{4\pi}\right)^2 C_F\Big(C_F-\frac{C_A}{2}\Big)
	\big( \mu^2 \bar \tau_L^2\big)^{2\eps}
	\big(\nu_+ \bar \tau_L\big)^{2\alpha}
\nonumber\\
   &\quad\times
   \bigg\{ \bigg[ \frac{4}{\eps^2} +\frac{16\ln z_+}{\eps}
   	   + 16 \dilog\Big(\! -\frac{z_-}{z_+} \Big) +32\ln^2 z_+ +2 \pi^2 	 
   	   \bigg]\frac{1}{\alpha^2}
\nonumber\\
   &\hspace{1.3cm}+ \bigg[ \!
	-\frac{4}{\eps^3}  -\frac{8\ln z_+}{\eps^2}
	+ \bigg( 8\dilog\Big(\! -\frac{z_-}{z_+} \Big) + \frac{2}{3}\pi^2 
	\bigg)\frac{1}{\eps}
	-48 \nielsen\Big(\! -\frac{z_-}{z_+} \Big) 
	+8 \trilog\Big(\! -\frac{z_-}{z_+} \Big)  
	\nonumber\\
   &\hspace{2.1cm}
	+64 \ln z_+ \,\dilog\Big(\! -\frac{z_-}{z_+} \Big)
	+\frac{64}{3} \ln^3 z_+
	+\frac{20\pi^2}{3} \ln z_+ + \frac{40}{3} \zeta_3 \bigg]\frac{1}{\alpha}\bigg\},
\end{align}
which is valid up to finite terms in the $\alpha$-expansion. Here and below, we drop the index $L$ on the variables $z^L_\pm = (\sqrt{1+z_L^2}\pm1)/4$ for convenience.

The calculation of the remaining diagrams proceeds along the same lines as in Section \ref{sec:twoparticlecut}. In broadening-momentum space, the non-abelian diagrams can be written in the form
\begin{align}
	&{\cal S}_L^{(2d/2e)}(b_L,b_R,p_L^\perp, p_R^\perp)
\\
   &\quad
= -\frac{\alpha_s^2 C_F C_A}{16\pi^{3-2\eps}}\, 
   \left(\frac{\mu^2e^{\gamma_E}}{b_L}\right)^{2\eps} \frac{\Omega_{d-3}}{b_L^3}\;\frac{1}{\alpha}\,
   \left(\frac{\nu_+}{b_L}\right)^{2\alpha} \!(1-y^2)^{-1-\eps} \,
    \delta^{d-2}(p_R^\perp)\;
    \delta(b_R)\,I_S^{(2d/2e)}(y)
\nonumber
\end{align}
with the two-dimensional integrals
\begin{align}
   I_S^{(2d)}(y) &= \frac{\sqrt{1-y^2}}{\pi} \;y^{2\eps}\!
   	\int_{1-y}^{1+y} \!\!d\xi\;\xi^{-1-2\alpha}\, (2-\xi)\,
	\int_0^{\frac{2-\xi}{\xi}} \!d\rho \;\frac{\rho^{-\alpha}}{1+\rho}
\nonumber\\
&\quad\times
	\bigg[ 1 - \frac{ \rho(3-\rho)\xi^2}{(2-\xi)^2+\rho\xi^2}\bigg]\;
	\frac{ (1+y-\xi)^{-\frac12-\eps}(\xi-1+y)^{-\frac12-\eps}}
	{(1+\rho)[(2-\xi)^2+\rho\xi^2]-4 y^2\rho}
\end{align}
and
\begin{align}
   I_S^{(2e)}(y) &= \frac{\sqrt{1-y^2}}{\pi} \;y^{2\eps}\!
   	\int_{1-y}^{1+y} \!\!d\xi\; \frac{\;\xi^{-1-2\alpha}\!\!\!}{2-\xi}\;
	\int_0^{\frac{2-\xi}{\xi}} \!d\rho \;\frac{\rho^{-1-\alpha}}{1+\rho}\;
	\frac{ (1+y-\xi)^{-\frac12-\eps}(\xi-1+y)^{-\frac12-\eps}}
	{(1+\rho)[(2-\xi)^2+\rho\xi^2]-4y^2\rho}
\nonumber\\
&\quad\times
	\bigg[ \xi^2 \rho^2 (1 + \rho) + (2 - \xi)^2 (1 + 2 \rho) 
	+ \frac{ \xi^2 (2 - \xi)^2 \rho^2 (1 + \rho)}{(2-\xi)^2+\rho\xi^2}\bigg]
		\,.
\end{align}
Note that the second integral generates a divergence in the analytic regulator in the limit $\rho\to 0$. Upon expansion in $\alpha$, we could solve both of these integrals in terms of hypergeometric functions, but the exact expressions are rather lengthy. In the limit $y\to1$, we find
\begin{align}
   I_S^{(2d)}(y\to 1) &\simeq 
	 2^{-1 - 2\eps}\;\frac{\Gamma(2 - \eps)\, \Gamma(- \eps)}{\Gamma(3 - 2\eps)}
	+(1-y)^{-\eps}\;2^{-1 - \eps}\;\frac{\Gamma(1 - 2 \eps)\, \Gamma(1 + 2 
	 \eps)}{\Gamma(1 - \eps)\,\Gamma(1 + \eps)}	
\nonumber\\
&\quad \times
	 \left\{   \frac{1}{2\eps}- \ln(1-y) - \ln2 + \Psi\Big(\frac12 + \eps\Big) - 
	\Psi(1 + \eps) \right\}
\end{align}
and
\begin{align}
   I_S^{(2e)}(y\to 1) &\simeq 
	 2^{-2 - 2\eps}\;\frac{3\Gamma(1 - \eps)\, \Gamma(- \eps)}{\Gamma(1 - 2\eps)}
	+(1-y)^{-\eps}\;2^{-1 - \eps}\;\frac{\Gamma(1 - 2 \eps)\, \Gamma(1 + 2 
	 \eps)}{\Gamma(1 - \eps)\,\Gamma(1 + \eps)}	
\nonumber\\
&\quad \times
	 \left\{   \frac{3}{2\eps}- \frac{2}{\alpha} + \ln(1-y) - 3\ln2 - \Psi\Big(\frac12 + \eps\Big) -3 \Psi(1 + \eps) - 4 \gamma_E \right\}\,.
\end{align}
This needs to be combined with the expansions for $y<1$, which are given by
\begin{align}
   &I_S^{(2d)}(y<1) = 	
	\frac{\sqrt{1-y^2}}{2y^3} \arcsin y - \frac{1}{y^2} \ln(1-y^2)
	- \frac{1}{4y^3} \ln \Big(\frac{1+y}{1-y}\Big) - \frac12
\nonumber\\
   &\quad	
	+ \frac{1}{y^2} \bigg\{ 2\dilog \Big( \frac{1-\sqrt{1-y^2}}{1+\sqrt{1-y^2}} \Big)
	- 2\dilog \Big( \frac{\sqrt{1-y^2}-1}{1+\sqrt{1-y^2}} \Big) 
	-\frac{2}{y} \,\dilog \Big( -\sqrt{\frac{1-y}{1+y}} \Big) 
	+\ln^2(1-y^2)
\nonumber\\
   &\hspace{1.6cm}
	+ \frac{\sqrt{1-y^2}}{2y} \bigg( 4 {\rm Cl}_2\big(\pi-\arcsin y\big)- {\rm 
	Cl}_2\big(\pi-2\arcsin y\big) \bigg)	
	-y^2 \ln(1+\sqrt{1-y^2}) 
\nonumber\\
   &\hspace{1.6cm}
   	-2\ln(1-y^2) \ln(1+\sqrt{1-y^2}) 
	+\frac{y^2}{2}\ln(1-y^2)
	-\frac{1+2y}{8y} \ln^2\Big(\frac{1-y}{1+y}\Big) 
\nonumber\\
   &\hspace{1.6cm}	
	-\frac{y^2+\ln2}{2y} \ln\Big(\frac{1-y}{1+y}\Big)
	-y^2 -\frac{\pi^2}{6y} \bigg\} \eps  
+ \calO(\eps^2)
\end{align}
and
\begin{align}
   &I_S^{(2e)}(y<1) = 	
   	-\frac{1}{\alpha}
   	\bigg\{ 1 + \bigg( 2 \ln(1+\sqrt{1-y^2}) - \ln(1-y^2) \bigg) \eps 
+ \bigg( 2\dilog \Big( \frac{1-\sqrt{1-y^2}}{1+\sqrt{1-y^2}} \Big)
\nonumber
\\
   &\quad
   	+ 2 \ln^2(1+\sqrt{1-y^2})
   	+\frac12 \ln^2(1-y^2)
   	-2\ln(1-y^2) \ln(1+\sqrt{1-y^2}) 
   	+\frac{\pi^2}{6} \bigg) \eps^2
   	+ \calO(\eps^3)\bigg\}
\nonumber\\
   &\quad   
	- 2 \ln\Big(\frac{1+\sqrt{1-y^2}}{2}\Big) 
	- \frac{1}{y^2} \ln(1-y^2)
	- \frac{3}{4y} \ln \Big(\frac{1+y}{1-y}\Big) 
\nonumber\\
   &\quad	
	+ \frac{1}{y^2} \bigg\{ 
	2(1-y^2)\dilog \Big( \frac{1-\sqrt{1-y^2}}{1+\sqrt{1-y^2}} \Big)
	- 2(1+2y^2)\dilog \Big( \frac{\sqrt{1-y^2}-1}{1+\sqrt{1-y^2}} \Big) 
	-2y^2\ln2\,\ln(1-y^2)
\nonumber\\
   &\hspace{1.6cm}
   	-6y \,\dilog \Big( -\sqrt{\frac{1-y}{1+y}} \Big) 
	-4y^2 \ln^2(1+\sqrt{1-y^2})
	-\frac{3y}{2} \ln2\, \ln\Big(\frac{1-y}{1+y}\Big) 
	+\ln^2(1-y^2)
\nonumber\\
   &\hspace{1.6cm}
   	-2(1-y^2)\ln(1-y^2) \ln(1+\sqrt{1-y^2}) 
	-\frac{2+3y+4y^2}{8} \ln^2\Big(\frac{1-y}{1+y}\Big) 
\nonumber\\
   &\hspace{1.6cm}	
	+4y^2\ln2\, \ln(1+\sqrt{1-y^2}) 	
	-\frac{y}{2}\pi^2
	 \bigg\} \eps  
+ \calO(\eps^2).
\end{align}
In Laplace-Fourier space the divergent terms in the $\alpha$-expansion become
\begin{align}
&\overline{\cal S}_L^{(2d)}(\tau_L,\tau_R,z_L, z_R) 
   =   
\left(\frac{\alpha_s}{4\pi}\right)^2 \frac{C_F C_A}{2}
	\big( \mu^2 \bar \tau_L^2\big)^{2\eps}
	\frac{\big(\nu_+ \bar \tau_L\big)^{2\alpha}}{\alpha}
	\nonumber\\
   &\quad\times\bigg\{
	\frac{1}{\eps^3}  +\frac{4\ln z_++2}{\eps^2}
	+ \bigg( 12\dilog\Big(\! -\frac{z_-}{z_+} \Big) 
	+8\ln^2 z_+ +8\ln z_+
	+ 4 +\frac76 \pi^2 \bigg)\frac{1}{\eps}
	+16\,h_2(z_L)
	\nonumber\\
   &\hspace{1.1cm}	
	-8 \nielsen\Big(\! -\frac{z_-}{z_+} \Big) 
	+4 \trilog\Big(\! -\frac{z_-}{z_+} \Big)
	+8 \nielsen(-w) 
	-40 \trilog(-w)
   	-40 \nielsen(1-w) 
	\nonumber\\
   &\hspace{1.1cm}	
	+8 \trilog(1-w)
   	+40 \nielsen\Big( \frac{1-w}{2} \Big) 
	-8 \trilog\Big( \frac{1-w}{2} \Big)
	+8 \ln \Big(\frac{z_+}{64}\Big) \,\dilog\Big(\! -\frac{z_-}{z_+} \Big)
	\nonumber\\
   &\hspace{1.1cm}	
   	-8 \ln2 \,\dilog\Big( \frac{1-w}{2} \Big)
	+\Big( 20 \ln(1+z_L^2) + 8 \ln(4z_+)\Big) \dilog(-w)
	+8 \dilog\Big(\! -\frac{z_-}{z_+} \Big)
	+\frac{8}{3} \ln^3 z_+ 
	\nonumber\\
   &\hspace{1.1cm}	
	+2\ln(1+z_L^2) \ln^2 z_+
	+5\ln^2(1+z_L^2) \ln z_+
	+(16-28\ln2)\ln^2 z_+
	+8\ln2\,\ln(1+z_L^2) \ln z_+
	\nonumber\\
   &\hspace{1.1cm}		
	+10\ln2\,\ln^2(1+z_L^2) 
	+8\Big(2 + \frac{\pi^2}{3} - 4 \ln^2 2-2w\Big) \ln z_+
	-16(1+2\ln2)w
	\nonumber\\
   &\hspace{1.1cm}
	+8\Big(\ln^2 2+w\Big) \ln (1+z_L^2)
	-12\ln^3 2+ 24+\pi^2 -\frac{4\pi^2}{3}\ln2-\frac{10}{3}\zeta_3
   \bigg\}
\end{align}
and
\begin{align}
&\overline{\cal S}_L^{(2e)}(\tau_L,\tau_R,z_L, z_R) 
   =   
\left(\frac{\alpha_s}{4\pi}\right)^2 \frac{C_F C_A}{2}
	\big( \mu^2 \bar \tau_L^2\big)^{2\eps}
	\big(\nu_+ \bar \tau_L\big)^{2\alpha}
\\
   &\quad\times
   \bigg\{ \bigg[ \frac{2}{\eps^2} +\frac{8\ln z_+}{\eps}
   	   + 8 \dilog\Big(\! -\frac{z_-}{z_+} \Big) +16\ln^2 z_+ +\pi^2 	 
   	   \bigg]\frac{1}{\alpha^2}
\nonumber\\
   &\hspace{1.3cm}+ \bigg[ 
	\frac{4\ln z_+}{\eps^2}
	+ \bigg( 28\dilog\Big(\! -\frac{z_-}{z_+} \Big) 
	+ 16\ln^2 z_++ 2\pi^2 \bigg)\frac{1}{\eps}
	-72 \nielsen\Big(\! -\frac{z_-}{z_+} \Big) 
	+20 \trilog\Big(\! -\frac{z_-}{z_+} \Big)  
	\nonumber\\
   &\hspace{2.0cm}	
	+24 \nielsen(-w)
	-56 \trilog(-w)
   	-56 \nielsen(1-w)
	+24 \trilog(1-w)
   	+56 \nielsen\Big( \frac{1-w}{2} \Big) 
	\nonumber\\
   &\hspace{2.0cm}	
	-24 \trilog\Big( \frac{1-w}{2} \Big)
	+72 \ln \Big(\frac{z_+}{2}\Big) \,\dilog\Big(\! -\frac{z_-}{z_+} \Big)
	-8\ln2 \,\dilog\Big(\! -\frac{z_-}{z_+} \Big)
	+\frac{56}{3} \ln^3 z_+ 
	\nonumber\\
   &\hspace{2.0cm}
	-24 \ln2 \,\dilog\Big( \frac{1-w}{2} \Big)
	+\Big( 28 \ln(1+z_L^2) + 24 \ln(4z_+)\Big) \dilog(-w)
	+6\ln(1+z_L^2) \ln^2 z_+
	\nonumber\\
   &\hspace{2.0cm}
	+7\ln^2(1+z_L^2) \ln z_+
	-52\ln2\,\ln^2 z_+
	+24\ln2\,\ln(1+z_L^2) \ln z_+
	+14\ln2\,\ln^2(1+z_L^2)
	\nonumber\\
   &\hspace{2.0cm}	
	-\Big(64 \ln^2 2-\frac{28}{3}\pi^2\Big) \ln z_+
	+24 \ln^2 2\, \ln (1+z_L^2)		 
	-\frac{76}{3}\ln^3 2+\frac{4\pi^2}{3}\ln2-6\zeta_3
	\bigg]\frac{1}{\alpha}\bigg\},
\nonumber
\end{align}
where $z_\pm = (w\pm1)/4$, $w=\sqrt{1+z_L^2}$ and the function $h_2(z)$ from (\ref{eq:def:h3}).

We next turn to the quark-loop contributions to the self-energy diagrams, which we write in the form
\begin{align}
	&{\cal S}_L^{(2f/2g,n_f)}(b_L,b_R,p_L^\perp, p_R^\perp)
\\
   &\quad
= -\frac{\alpha_s^2 C_F T_F n_f}{4\pi^{3-2\eps}}\, 
   \left(\frac{\mu^2e^{\gamma_E}}{b_L}\right)^{2\eps} \frac{\Omega_{d-3}}{b_L^3}\;\frac{1}{\alpha}\,
   \left(\frac{\nu_+}{b_L}\right)^{2\alpha} \!(1-y^2)^{-2-\eps} \,
    \delta^{d-2}(p_R^\perp)\;
    \delta(b_R)\,I_S^{(2f/2g,n_f)}(y)\,.
    \nonumber
\end{align}
This time, we factored out the distribution $(1-y^2)^{-2-\eps}$. The two-dimensional integral representations of the diagrams read
\begin{align}
   I_S^{(2f,n_f)}(y) &= 4y^{2\eps}\,\frac{(1-y^2)^{3/2}}{\pi} \!
   	\int_{1-y}^{1+y} \!\!d\xi\;\xi^{1-2\alpha}\, (2-\xi)\,
	\int_0^{\frac{2-\xi}{\xi}} \!d\rho \;\frac{\rho^{2-\alpha}}{1+\rho}
\nonumber\\
&\quad\times
	\frac{ 2(1-y^2)-\xi(2-\xi)}{[(2-\xi)^2+\rho\xi^2]}\;\;
	\frac{ (1+y-\xi)^{-\frac12-\eps}(\xi-1+y)^{-\frac12-\eps}}
	{[(1+\rho)[(2-\xi)^2+\rho\xi^2]-4 y^2\rho]^2}
\end{align}
and
\begin{align}
   I_S^{(2g,n_f)}(y) &= y^{2\eps}\,\frac{(1-y^2)^{3/2}}{\pi} \!
   	\int_{1-y}^{1+y} \!\!d\xi\;\xi^{1-2\alpha}\, (2-\xi)\,
	\int_0^{\frac{2-\xi}{\xi}} \!d\rho \;\frac{\rho^{2-\alpha}}{(1+\rho)^2}
\nonumber\\
&\quad\times
	\bigg[ 1 + \frac{ \xi^2(2-\xi)^2(1+\rho)^2}{[(2-\xi)^2+\rho\xi^2]^2}\bigg]\;\,
	\frac{ (1+y-\xi)^{-\frac12-\eps}(\xi-1+y)^{-\frac12-\eps}}
	{[(1+\rho)[(2-\xi)^2+\rho\xi^2]-4 y^2\rho]^2}\,.
\end{align}
As we factored out the distribution $(1-y^2)^{-2-\eps}$, we now have to expand to subleading order around the limit $y\to1$. We find
\begin{align}
   I_S^{(2f,n_f)}(y\to 1) &\simeq 
	 -2^{-1 - 2\eps}\;\frac{\Gamma(2 - \eps)\, \Gamma(3- \eps)}{\Gamma(5 - 2\eps)}
	 \bigg[ 1 - \frac{8-7\eps}{1-\eps} \,(1-y)\bigg]
\end{align}
and 
\begin{align}
   I_S^{(2g,n_f)}(y\to 1) &\simeq 
	 2^{-2 - 2\eps}\;\frac{\Gamma(2 - \eps)\, \Gamma(3- \eps)}{\Gamma(5 - 2\eps)}
	 \bigg[ 1 + \frac{3\eps}{1-\eps} \,(1-y)\bigg].
\end{align}
For arbitrary $y<1$ the expansions become
\begin{align}
   &I_S^{(2f,n_f)}(y<1) = 	
	\frac{1-3y^2+2y^4}{8y^5} \sqrt{1-y^2}\arcsin y 
	+ \frac{(5-7y^2)(1-y^2)^2}{64y^5} \ln \Big(\frac{1-y}{1+y}\Big) 
\nonumber\\
   &\quad
	+ \frac{3-6y^2+15y^4-16y^6}{96y^4}	
	+ \bigg\{ \frac{3-9y^2+37y^4-39y^6}{192y^5} \ln\Big(\frac{1-y}{1+y}\Big) 
	+\frac{3-6y^2+7y^4}{48y^4} \sqrt{1-y^2}
\nonumber\\
   &\quad
	+ \frac{1-3y^2+2y^4}{8y^5} \sqrt{1-y^2}\bigg( 4 {\rm Cl}_2\big(\pi-\arcsin y\big)- {\rm 
	Cl}_2\big(\pi-2\arcsin y\big) \bigg)
\nonumber\\
   &\quad
   	- \frac{(5-7y^2)(1-y^2)^2}{128y^5}\bigg( 
	16\dilog \Big( -\sqrt{\frac{1-y}{1+y}} \Big)
	+ \ln^2\Big(\frac{1-y}{1+y}\Big)
	+ 4\ln2\,\ln\Big(\frac{1-y}{1+y}\Big)
	+\frac{4}{3}\pi^2\bigg)
\nonumber\\
   &\quad
	+\frac{3-6y^2+15y^4-16y^6}{48y^4} \ln\Big(\frac{1+\sqrt{1-y^2}}{\sqrt{1-y^2}} \Big)
	-\frac{9-18y^2-51y^4+80y^6}{288y^4}
	  \bigg\} \eps  
+ \calO(\eps^2)
\end{align}
and
\begin{align}
   &I_S^{(2g,n_f)}(y<1) = 	
	-\frac{3(3-y^2)(1-y^2)^2}{128y^5} \ln \Big(\frac{1-y}{1+y}\Big) 
	- \frac{27-54y^2+23y^4}{192y^4}
\nonumber\\
   &\quad	
	+ \bigg\{ 
	- \frac{27-54y^2+23y^4}{96y^4} \ln\Big(\frac{1+\sqrt{1-y^2}}{\sqrt{1-y^2}} \Big)	
	- \frac{99-233y^2+165y^4-39y^6}{384y^5} \ln\Big(\frac{1-y}{1+y}\Big) 
\nonumber\\
   &\hspace{1.2cm}
   	+\frac{3(3-y^2)(1-y^2)^2}{256y^5}\bigg( 
	16\dilog \Big( -\sqrt{\frac{1-y}{1+y}} \Big)
	+ \ln^2\Big(\frac{1-y}{1+y}\Big)
	+ 4\ln2\,\ln\Big(\frac{1-y}{1+y}\Big)
	+\frac{4}{3}\pi^2\bigg)
\nonumber\\
   &\hspace{1.2cm}	
	-\frac{135-294y^2+139y^4}{576y^4}
	-\frac{27-42y^2+19y^4}{96y^4} \sqrt{1-y^2}  \bigg\} \eps  
+ \calO(\eps^2)\,.
\end{align}
In Laplace-Fourier space this finally translates into
\begin{align}
&\overline{\cal S}_L^{(2f,n_f)}(\tau_L,\tau_R,z_L, z_R) 
   =   
\left(\frac{\alpha_s}{4\pi}\right)^2 C_F T_F n_f\;
	\big( \mu^2 \bar \tau_L^2\big)^{2\eps}\,
	\frac{\big(\nu_+ \bar \tau_L\big)^{2\alpha}}{\alpha}
	\\
   &\quad\times\bigg\{\!
	- \frac{4}{3\eps^2}  
	- \Big( \frac{16}{3} \ln z_+ -\frac{2}{3w} + \frac{20}{9} \Big) \frac{1}{\eps}
	+\frac89 (1+z_L^2) \,h_1(z_L)
	-\frac89 (13+2z_L^2) \,h_2(z_L)
	\nonumber\\
   &\hspace{1.2cm}	
	-\frac{16}{3} \dilog\Big(\! -\frac{z_-}{z_+} \Big)
	-\frac{32}{3} \ln^2 z_+ 
	-\frac{80}{9} \ln z_+ 
	+ \frac{116+20z_L^2}{9}\,w\ln z_+
	-\frac{20}{9} z_L^2 
	-\frac{472}{27}
	-\frac{2}{3}\pi^2
	\nonumber\\
   &\hspace{1.2cm}
   	-\frac{46+68z_L^2+10z_L^4}{9w}\ln(1+z_L^2)
   	+\frac{16+40\ln2}{9w}z_L^4
   	+\frac{122+272\ln2}{9w}z_L^2
   	+\frac{110+196\ln2}{9w}
   \bigg\}
  \nonumber
\end{align}
and
\begin{align}
&\overline{\cal S}_L^{(2g,n_f)}(\tau_L,\tau_R,z_L, z_R) 
   =   
\left(\frac{\alpha_s}{4\pi}\right)^2 C_F T_F n_f\;
	\big( \mu^2 \bar \tau_L^2\big)^{2\eps}\,
	\frac{\big(\nu_+ \bar \tau_L\big)^{2\alpha}}{\alpha}
	\;\frac{1}{w}
	\\
   &
   	\bigg\{\!
	- \frac{1}{3\eps}  
	- 2(1+z_L^2)^2 \ln(1+w)
	+\frac{1+6z_L^2+3z_L^4}{3} \ln(1+z_L^2)
	- z_L^2 + 2\ln2 - \frac{11}{9}
	+ \Big( \frac83 + 2 z_L^2 \Big) w
   \bigg\},
  \nonumber
\end{align}
where the functions $h_1(z)$ and $h_2(z)$ can be found in (\ref{eq:def:h3}).

We proceed similarly for the gluon and ghost contributions to the self-energy diagrams, which we add up and write in the form
\begin{align}
	&{\cal S}_L^{(2f/2g,C_A)}(b_L,b_R,p_L^\perp, p_R^\perp)
\\
   &\quad
= \frac{\alpha_s^2 C_F C_A}{32\pi^{3-2\eps}}\, 
   \left(\frac{\mu^2e^{\gamma_E}}{b_L}\right)^{2\eps} \frac{\Omega_{d-3}}{b_L^3}\;\frac{1}{\alpha}\,
   \left(\frac{\nu_+}{b_L}\right)^{2\alpha} \!(1-y^2)^{-2-\eps} \,
    \delta^{d-2}(p_R^\perp)\;
    \delta(b_R)\,I_S^{(2f/2g,C_A)}(y)
\nonumber
\end{align}
with
\begin{align}
   &I_S^{(2f,C_A)}(y) = 4y^{2\eps}\,\frac{(1-y^2)^{3/2}}{\pi} \!
   	\int_{1-y}^{1+y} \!\!d\xi\;\xi^{1-2\alpha}\, (2-\xi)\,
	(1+y-\xi)^{-\frac12-\eps}(\xi-1+y)^{-\frac12-\eps} 
\\
&\quad\times
\int_0^{\frac{2-\xi}{\xi}} \!d\rho\;
\frac{\rho^{1-\alpha}}{1+\rho}\;
	\frac{ (2-\xi)^2+\rho^2\xi^2-6\rho(2-\xi)\xi+12\rho-16y^2\rho+\eps(1-\rho)[(2-\xi)^2-\rho\xi^2]}{[(2-\xi)^2+\rho\xi^2]\;[(1+\rho)[(2-\xi)^2+\rho\xi^2]-4 y^2\rho]^2}
	\nonumber
\end{align}
and
\begin{align}
   &I_S^{(2g,C_A)}(y) = y^{2\eps}\,\frac{(1-y^2)^{3/2}}{\pi} \!
   	\int_{1-y}^{1+y} \!\!d\xi\;\xi^{1-2\alpha}\, (2-\xi)\,
	\int_0^{\frac{2-\xi}{\xi}} \!d\rho \;
	\frac{ (1+y-\xi)^{-\frac12-\eps}(\xi-1+y)^{-\frac12-\eps}}
	{[(1+\rho)[(2-\xi)^2+\rho\xi^2]-4 y^2\rho]^2}
\nonumber\\
&\quad\times
	\rho^{1-\alpha}\;\bigg[ \frac{ (1+\eps)(1+\rho^2)+(6-2\eps)\rho }{(1+\rho)^2} 
	+ \frac{ (1+\eps)[(2-\xi)^4+\rho^2\xi^4]+(6-2\eps)\rho\xi^2(2-\xi)^2}{[(2-\xi)^2+\rho\xi^2]^2}\bigg]
\,.
\end{align}
The solutions in the limit $y\to1$ now read
\begin{align}
   I_S^{(2f,C_A)}(y\to 1) &\simeq 
	 -2^{- 2\eps}\;\frac{\Gamma(1 - \eps)\, \Gamma(3- \eps)}{\Gamma(5 - 2\eps)}
	 \Big[ 5-3\eps - (43 - 21 \eps - 4 \eps^2) \,(1-y)\Big]
\nonumber\\
&\quad
	 -2^{-1- \eps}\;(1-y)^{1-\eps}\,\frac{\Gamma(1 - 2\eps)\, \Gamma(1+2 \eps)}
	 {\Gamma(2-\eps)\Gamma(1+\eps)}
\end{align}
and 
\begin{align}
   I_S^{(2g,C_A)}(y\to 1) &\simeq 
	 2^{-1- 2\eps}\;\frac{\Gamma(1 - \eps)\, \Gamma(3- \eps)}{\Gamma(5 - 2\eps)}
	 \Big[ 5-3\eps - (3-5 \eps + 4 \eps^2) \,(1-y)\Big]
\nonumber\\
&\quad
	 +2^{-2- \eps}\;(1-y)^{1-\eps}\,\frac{\Gamma(1 - 2\eps)\, \Gamma(1+2 \eps)}
	 {\Gamma(2-\eps)\Gamma(1+\eps)}\,,
\end{align}
and the expansions for $y<1$ are
\begin{align}
   &I_S^{(2f,C_A)}(y<1) = 	
	\frac{(1-4y^2)(1-y^2)^{3/2}}{2y^5} \arcsin y 
	+ \frac{(5-16y^2)(1-y^2)^2}{16y^5} \ln \Big(\frac{1-y}{1+y}\Big) 
\nonumber\\
   &\quad
	+ \frac{3-9y^2+36y^4-40y^6}{24y^4}	
	+ \bigg\{ \frac{3-9y^2+16y^4}{12y^4} \sqrt{1-y^2}
	- \frac{3-3y^2-22y^4+27y^6}{12y^5} \ln\Big(\frac{1-y}{1+y}\Big) 
\nonumber\\
   &\quad
	+ \frac{(1-4y^2)(1-y^2)^{3/2}}{2y^5}\bigg( 4 {\rm Cl}_2\big(\pi-\arcsin y\big)- {\rm 
	Cl}_2\big(\pi-2\arcsin y\big) \bigg)
	- \frac{(1-y^2)^{3/2}}{2y^5} \arcsin y
\nonumber\\
   &\quad
   	- \frac{(5-16y^2)(1-y^2)^2}{32y^5}\bigg( 
	16\dilog \Big( -\sqrt{\frac{1-y}{1+y}} \Big)
	+ \ln^2\Big(\frac{1-y}{1+y}\Big)
	+ 4\ln2\,\ln\Big(\frac{1-y}{1+y}\Big)
	+\frac{4}{3}\pi^2\bigg)
\nonumber\\
   &\quad	
	+\frac{3-9y^2+36y^4-40y^6}{12y^4} \ln\Big(\frac{1+\sqrt{1-y^2}}{\sqrt{1-y^2}} \Big)
	-\frac{9-18y^2-84y^4+124y^6}{36y^4}
	  \bigg\} \eps  
+ \calO(\eps^2)
\end{align}
and
\begin{align}
   &I_S^{(2g,C_A)}(y<1) = 	
	-\frac{(9-4y^2)(1-y^2)^2}{32y^5} \ln \Big(\frac{1-y}{1+y}\Big) 
	- \frac{(9-4y^2)(3-5y^2)}{48y^4}
\nonumber\\
   &\quad	
	+ \bigg\{ 
	- \frac{(9-4y^2)(3-5y^2)}{24y^4} \ln\Big(\frac{1+\sqrt{1-y^2}}{\sqrt{1-y^2}} \Big)	
	- \frac{18-44y^2+33y^4-12y^6}{24y^5} \ln\Big(\frac{1-y}{1+y}\Big) 
\nonumber\\
   &\quad
   	+\frac{(9-4y^2)(1-y^2)^2}{64y^5}\bigg( 
	16\dilog \Big( -\sqrt{\frac{1-y}{1+y}} \Big)
	+ \ln^2\Big(\frac{1-y}{1+y}\Big)
	+ 4\ln2\,\ln\Big(\frac{1-y}{1+y}\Big)
	+\frac{4}{3}\pi^2\bigg)
\nonumber\\
   &\quad	
	-\frac{27-66y^2+8y^4}{72y^4}
	-\frac{27-45y^2+28y^4}{24y^4} \sqrt{1-y^2}  \bigg\} \eps  
+ \calO(\eps^2)\,.
\end{align}
In Laplace-Fourier space we finally obtain
\begin{align}
&\overline{\cal S}_L^{(2f,C_A)}(\tau_L,\tau_R,z_L, z_R) 
   =   
\left(\frac{\alpha_s}{4\pi}\right)^2 \frac{C_F C_A}{2}\;
	\big( \mu^2 \bar \tau_L^2\big)^{2\eps}\,
	\frac{\big(\nu_+ \bar \tau_L\big)^{2\alpha}}{\alpha}
	\\
   &\quad\times\bigg\{\!
	\frac{10}{3\eps^2}  
	+ \Big( \frac{40}{3} \ln z_+ -\frac{5}{3w} + \frac{62}{9} \Big) \frac{1}{\eps}
	-\frac89 (1+z_L^2) \,h_1(z_L)
	+\frac89 (31+2z_L^2) \,h_2(z_L)
	+\frac{5}{3}\pi^2
	\nonumber\\
   &\hspace{1.2cm}	
	+\frac{40}{3} \dilog\Big(\! -\frac{z_-}{z_+} \Big)
	+\frac{80}{3} \ln^2 z_+ 
	+\frac{248}{9} \ln z_+ 
	- \frac{278+20z_L^2}{9}\,w\ln z_+
   	-\frac{266+596\ln2}{9w}z_L^2
	\nonumber\\
   &\hspace{1.2cm}
   	+\frac{109+149z_L^2+10z_L^4}{9w}\ln(1+z_L^2)
   	-\frac{16+40\ln2}{9w}z_L^4
   	-\frac{266+466\ln2}{9w}
	+\frac{20}{9} z_L^2 
	+\frac{1222}{27}
   \bigg\}
  \nonumber
\end{align}
and
\begin{align}
&\overline{\cal S}_L^{(2g,C_A)}(\tau_L,\tau_R,z_L, z_R) 
   =   
\left(\frac{\alpha_s}{4\pi}\right)^2 \frac{C_F C_A}{2}\;
	\big( \mu^2 \bar \tau_L^2\big)^{2\eps}\,
	\frac{\big(\nu_+ \bar \tau_L\big)^{2\alpha}}{\alpha}
	\;\frac{1}{w}
  \nonumber	\\
   &\quad\times
   	\bigg\{\!
	\frac{5}{6\eps}  
	+(1+z_L^2)(3+2z_L^2) \ln(1+w)
	+\frac{1-15z_L^2-6z_L^4}{6} \ln(1+z_L^2)
	\nonumber\\
   &\hspace{1.2cm}
	+ z_L^2 -5\ln2 + \frac{17}{9}
	- \Big( \frac{11}{3} + 2 z_L^2 \Big) w
   \bigg\}.
\end{align}

\end{appendix}

\end{document}